\documentclass[fleqn,usenatbib]{mnras}

\usepackage{newtxtext,newtxmath}

\usepackage[T1]{fontenc}

\DeclareRobustCommand{\VAN}[3]{#2}
\let\VANthebibliography\thebibliography
\def\thebibliography{\DeclareRobustCommand{\VAN}[3]{##3}\VANthebibliography}

\usepackage{graphicx,subcaption}	
\usepackage{amsmath}	
\usepackage{xcolor}
\usepackage{float}
\usepackage{dcolumn}
\usepackage{sidecap}
\usepackage{multirow}
\usepackage[flushleft]{threeparttable}
\usepackage[hypcap=false]{caption}
\usepackage{copyrightbox}
\usepackage{listings}
\usepackage{multicol} 
\usepackage{afterpage}

\usepackage{enumitem}

\DeclareCaptionFormat{myformat}{\textbf{#1#2}#3\par}
\definecolor{lightgray}{rgb}{0.95, 0.95, 0.95}

\newcolumntype{L}{D{.}{.}{2,5}}

\newcommand{\astronomaly}{\textsc{astronomaly}}
\newcommand{\glade}{\citep[GLADE,][]{GLADE2018}}
\newcommand{\nvss}{\citep[NVSS,][]{1998AJ....115.1693C}}
\newcommand{\galex}{\citep[GALEX,][]{2005ApJ...619L...1M}}
\newcommand{\twomass}{\citep[2MASS,][]{2006AJ....131.1163S}}
\newcommand{\sixdf}{\citep[6dF][]{6df2009}}

\newcommand{\email}[1]{\href{mailto:#1}{#1}}

\usepackage{array}
\newcommand\Tstrut{\rule{0pt}{2.6ex}}         

\title[Astronomaly at scale]{Astronomaly at scale: searching for anomalies amongst 4 million galaxies}

\author[V. Etsebeth et al.]{
V. Etsebeth,$^{1}$\thanks{E-mail: \email{verlon18@gmail.com}} M. Lochner$^{1,2}$, M. Walmsley$^{3,4}$, M. Grespan$^{5}$
\\
$^{1}$Department of Physics and Astronomy, University of the Western Cape, Bellville, Cape Town, 7535, South Africa \\
$^{2}$South African Radio Astronomy Observatory, 2 Fir Street, Black River Park, Observatory 7925, South Africa \\
$^{3}$Jodrell Bank Centre for Astrophysics, Department of Physics \& Astronomy, University of Manchester, Oxford Road, Manchester M13 9PL, UK \\
$^{4}$Dunlap Institute for Astronomy \& Astrophysics, University of Toronto, 50 St George Street, Toronto, ON M5S 3H4, Canada \\
$^{5}$National Center for Nuclear Research, Andrzeja Sołtana 7/3, PL05-400 Otwock, Poland
}

\date{Accepted 2024 January 30. Received 2024 January 11; in original form 2023 October 4}

\pubyear{2024}

\begin{document}
\label{firstpage}
\pagerange{\pageref{firstpage}--\pageref{lastpage}}
\maketitle

\begin{abstract}
Modern astronomical surveys are producing data sets of unprecedented size and richness, increasing the potential for high-impact scientific discovery. This possibility, coupled with the challenge of exploring a large number of sources, has led to the development of novel machine-learning-based anomaly detection approaches, such as \astronomaly{}. For the first time, we test the scalability of \astronomaly{} by applying it to almost 4 million images of galaxies from the Dark Energy Camera Legacy Survey. We use a trained deep learning algorithm to learn useful representations of the images and pass these to the anomaly detection algorithm isolation forest, coupled with \astronomaly{}'s active learning method, to discover interesting sources. We find that data selection criteria have a significant impact on the trade-off between finding rare sources such as strong lenses and introducing artefacts into the data set. We demonstrate that active learning is required to identify the most interesting sources and reduce artefacts, while anomaly detection methods alone are insufficient. Using \astronomaly{}, we find 1635 anomalies among the top 2000 sources in the data set after applying active learning, including eight strong gravitational lens candidates, 1609 galaxy merger candidates, and 18 previously unidentified sources exhibiting highly unusual morphology. Our results show that by leveraging the human-machine interface, \astronomaly{} is able to rapidly identify sources of scientific interest even in large data sets.
\end{abstract}

\begin{keywords}
gravitational lensing: strong --
methods: data analysis -- 
surveys -- 
galaxies: general -- 
galaxies: interactions. \vspace{-6pt}
\end{keywords}


\section{Introduction}
\label{Introduction}
Astronomy, like many other fields, is experiencing a data revolution with large and rich data sets being produced at an extraordinary rate. Modern surveys, such as the Sloan Digital Sky Survey (SDSS; York et al. \citeyear{2000AJ....120.1579Y}), have already observed and classified over a billion objects across one-third of the sky  (Almeida et al. \citeyear{almeida2023eighteenth}). The Dark Energy Spectroscopic Instrument (DESI; Dey et al. \citeyear{Dey_2019}) Legacy Surveys cover a smaller area of the sky than the SDSS but obtained deeper and higher quality images and have observed 1.6 billion sources, including stars, galaxies, and quasars (Dey et al. \citeyear{Dey_2019}). Upcoming surveys from the Vera C. Rubin Observatory\footnote{\url{https://www.lsst.org/}} and the Square Kilometre Array\footnote{\url{https://www.skao.int/}} are expected to acquire petabytes of data and observe unprecedented numbers of astronomical phenomena. It is expected that the Vera C. Rubin Observatory will provide 32 trillion observations of 20 billion galaxies at a greater depth than previous large surveys  (Ivezić et al. \citeyear {2019ApJ...873..111I}). In addition, Euclid is poised to contribute significantly to the wealth of astronomical data (Scaramella et al. \citeyear{Euclid_2022}) and aims to map the geometry of the dark Universe with unprecedented accuracy (Tutusaus, Sorce \& Troja \citeyear{tutusaus2023euclid}).

Improved data quality and deeper images provide a better opportunity to detect rare and unknown astrophysical phenomena, called anomalies. Anomaly detection is often key for scientific discoveries but faces many challenges. For instance, a natural approach to look for rare objects might be to simulate data sets, as Metcalf et al. (\citeyear{Metcalf_2019}) did. However, Ciprijanovic et al. \citeyearpar{10.1093/mnras/stab1677} showed that this can actually fail when applied to real data sets as simulated and observational data often represent different data domains. 

Anomalies may already exist in data sets but have been previously overlooked, as seen when Massey, Neugent \& Levesque \citeyearpar{Massey_2019} identified quasi-stellar objects in various spectroscopic data sets. Large amounts of data make it more difficult to detect anomalies as manual inspections are not possible. Additionally, there are various types of anomalies, not all of which are scientifically interesting. Citizen science projects like Galaxy Zoo (Lintott et al. \citeyear{lintott2008}, \citeyear{lintott2011}) have been used to accelerate data investigation. The project's many volunteers perform critical galaxy classification and anomaly detection tasks, but they may overlook some features and might not necessarily have the training to identify interesting anomalies. 

Automated processes, such as machine learning (ML), have demonstrated their ability to handle complex tasks that were previously performed only by people. Supervised ML algorithms have been widely used in astronomy for classification tasks 
(e.g. Debosscher et al. \citeyear{Debosscher_2007}; Martinazzo, Espadoto \& Hirata \citeyear{cscv}; Soroka, Meshcheryakov \& Gerasimov \citeyear{MCAI}). While supervised methods fall short as true anomaly detectors, relying on training data and classifying predefined classes, unsupervised ML excels in anomaly detection by operating without the need for prior examples of the sources of interest. In astronomy, unsupervised ML has been applied to various data types. For instance, it has been used to identify outliers in SDSS spectroscopic data through adaptations of techniques like random forests (Baron \& Poznanski \citeyear{2017MNRAS.465.4530B}). Similarly, anomalies in Kepler light curves have been successfully detected (Giles \& Walkowicz \citeyear{2019MNRAS.484..834G}), and generative adversarial networks have been employed for anomaly detection in optical images (Storey-Fisher et al. \citeyear{Storey_Fisher_2021}). Solarz et al. \citeyearpar{Solarz_2017} showed that anomalies vary in relevance depending on the specific study. Anomalies may range from artefacts, which are of interest to system operators but considered contaminants by astronomers, to rare but known sources such as strong gravitational lenses, which have an estimated occurrence rate of just 1 in 10$^4$ (Huang et al. \citeyear{Huang_2021}). 

Unsupervised ML anomaly detection methods, when combined with additional techniques, have proven successful in astronomy. Lochner \& Bassett \citeyearpar{Lochner_2021} developed a general framework for anomaly detection that incorporates a novel active learning (AL) approach. \astronomaly{} combines AL with personalized user feedback, enabling users to interactively label objects and refine the anomaly detection process. Moreover, \astronomaly{} has the capability to handle various types of data, including images, spectra, or time series data, and can leverage domain knowledge and user preferences to improve detection accuracy and efficiency. \astronomaly{} was applied to the Galaxy Zoo\footnote{\url{https://www.kaggle.com/c/galaxy-zoo-the-galaxy-challenge}} data set, as well as on simulated data in order to evaluate the performance of the AL technique. In both cases, \astronomaly{}'s AL approach nearly doubled the detection of interesting anomalies in the initial 100 user-viewed objects compared to the popular anomaly detection algorithm, iForest (Liu, Ting \& Zhou \citeyear{10.1109/ICDM.2008.17}).
Walmsley et al. \citeyearpar{Walmsley_2022} adapted \astronomaly{} to include a deep learning approach combined with a novel AL algorithm. A convolutional neural network (CNN) was used to learn a low-dimensional representation that captures the salient features of galaxy images. Additionally, the regressor that models user interest: Random Forest (Breiman \citeyear{Breiman2001}) in \astronomaly{} was replaced by a Gaussian process (GP, Rasmussen \& Williams \citeyear{rasmussen2006gaussian}) that allowed the use of an acquisition function to more optimally select targets for user labelling. AHUNT, as an alternative approach, refines deep features through AL in each iteration to enhance anomaly detection (Vafaei Sadr, Bassett \& Sekyi \citeyear{sadr2022}).

Anomaly detection is a challenging task with very few studies done on a large scale in astronomy. 
While \astronomaly{} has been shown to be effective in detecting anomalies, it has also yet to be used on a large data set. The main objective of this work is to test the capabilities and limitations of \astronomaly{} by applying it to a large subset of the Dark Energy Camera Legacy Survey (DECaLS, Dey et al. \citeyear{Dey_2019}). DECaLS has yet to be extensively studied, making it excellent for searching for undiscovered anomalies. This will evaluate the performance and scalability of \astronomaly{} as well as provide the opportunity to make new discoveries.

The paper is structured as follows: Section \ref{Data} covers DECaLS subset selection criteria and data pre-processing. An evaluation set is also created, which is used to test the performance of the different AL methods mentioned previously. Section \ref{Methodology} discusses the \astronomaly{} framework, including the algorithms and parameters used for this work. Section \ref{Results: Evaluation Set} presents the findings from the evaluation set, followed by the results for the main DECaLS subset in Section \ref{Results: Application on a large scale}. Lastly, Section \ref{Results: Follow-up investigations} highlights some of the interesting anomalies detected.

\vspace{-10pt}

\section{Data}
\label{Data}
The DESI Legacy Surveys encompass three distinct imaging surveys: the Beijing--Arizona Sky Survey (BASS)\footnote{\url{https://batc.bao.ac.cn/BASS/doku.php}}, the Mayall $z$-band Legacy Survey\footnote{\url{https://www.legacysurvey.org/mzls/}}, and the DECaLS. These surveys are enriched with data from the Dark Energy Survey (The Dark Energy Survey Collaboration \citeyear{https://doi.org/10.48550/arxiv.astro-ph/0510346}) and select observations from the Wide-Field Infrared Survey Explorer (Wright et al. \citeyear{Wright_2010}), notably the W1 and W2 bands, to provide additional colour information. This work uses data from the eighth Public Data Release (DR8)\footnote{\url{https://www.legacysurvey.org/dr8/}} of DECaLS, using the three optical bands $g$, $r$, and $z$. DECalS reaches depths of magnitude 24, 23.4, and 22.5 in bands $g$, $r$, and $z$, respectively, which is significantly deeper than the previous large optical survey SDSS.

\subsection{Selection cuts}
\label{Data: Selection Cuts}
DR8 of the Legacy Surveys contains more than 1.6 billion unique sources. The quantity of data alone requires a subset to be selected due to limitations associated with downloading and storing such significant amounts of data. More information on the sources, storage, and computational requirements can be found in Appendix \ref{Appendix: Computational Resources}. This is a common challenge and colour or magnitude constraints are often implemented on target data (Sridhar et al. \citeyear{Sridhar_2020}; Walmsley et al. \citeyear{Walmsley_2022}). DECaLS also contains a number of sources that clearly do not have anomalous morphologies, such as stars, which should ideally be removed before searching for anomalies. While applying strict selection cuts can result in higher data quality and reduce the impact of artefacts and biases, it can also limit the scope of the analysis and remove potentially interesting anomalies. In this paper, we minimize the cuts used in order to produce a subset that is as inclusive as possible, but still of a manageable size for the resources available.

In order to implement appropriate selection cuts, we use the processing that has already been applied to DECaLS data to identify artefacts and classify sources, as outlined in Dey et al. \citeyearpar{Dey_2019}. It is important to acknowledge that the calibrations and techniques employed were not perfect. One area where challenges persist is in the detection and masking of artefacts and other bright sources. While the flagging from Dey et al. \citeyearpar{Dey_2019} allowed the removal of the bulk of these sources, a number of artefacts remained, which will be discussed in later sections. Here we briefly outline the final selection cuts used, which are described in more detail in Appendix \ref{Appendix: Data Cuts}.

First, we required no masked sources in any bands, which are generally masked due to artefacts or bright foreground stars.

Secondly, we ensure all sources are well fitted by a standard galaxy profile model. The DECaLS pipeline fits several models, including an exponential and de Vaucouleurs \citeyearpar{1948AnAp...11..247D} model, reporting on the reduced chi-square for the best-fitting model. We found that some sources had a reduced chi-square lower than one in all bands, which corresponded to artefacts, nearby masked sources, as well as faint and unresolved sources. We thus required all sources to have a reduced chi-square greater than one in at least one band. No upper limit was applied to the reduced chi-square as this would eliminate sources with unusual morphology which are anomalous but poorly fit by the profiles used.

A positive signal-to-noise (SNR) ratio in all three bands was also imposed. This cut also eliminated the observations that do not have passes in all three bands. We also applied a minimum flux threshold in any of the bands to remove sources that are too faint to be distinguishable. The threshold value was based on visual inspections and was chosen to ensure that only relatively bright sources were included. Sources were also selected based on their size, specifically whether they have a large enough radius in either of the two DECaLS models available (de Vaucouleurs or exponential). This cut reduced the number of compact objects that are unlikely to be resolved, focusing on more extended sources. These cuts were carefully chosen after testing various combinations of them and other criteria that are not listed in the final selection. For a more detailed description of the selection criteria, including the exact query used, see Appendix \ref{Appendix: Data Cuts}.

Mao et al. \citeyearpar{Mao_2021} imposed additional cuts by restricting the model fitting the parameters rchi\_g, rchi\_r, and rchi\_z, but it was seen that this removed a significant number of interesting sources when we applied this to a small subset of data, so they are not implemented. The DECaLS catalogue consists of  871 359 667 resolved (non-PSF) sources, which was reduced to 3 884 404 after applying the above selection cuts. This subset will be referred to as the main subset for the remainder of the paper.

\subsection{Image cut-out sizes}
\label{Data: Image Cutout Sizes}
Our methodology includes using a CNN to extract good representations of the source images as was done in Walmsley et al. \citeyearpar{Walmsley_2022}. CNNs generally require input images of identical dimensions, which makes choosing a cut-out size important. Moreover, since ML algorithms are affected by the angular source size in the image (Slijepcevic et al. \citeyear{slijepcevic2023radio}), we also needed to individually adjust the scale of the input images such that the sources are similarly sized. 

Walmsley et al. \citeyearpar{Walmsley_2022} employed a fixed image size that was possible because of the availability of the Petrosian radius (Petrosian \citeyear{1976ApJ...209L...1P}) for the sources, which allowed the pixel count that covered the target source to be estimated. Images were obtained from the DECaLS cut-out service, using the native telescope resolution and adjusting the visible sky area to match the radius determined by the Petrosian radii. Finally, the images in Walmsley et al. \citeyearpar{Walmsley_2022} were interpolated to 424x424 pixels and colourized for viewing purposes. 
However, the Petrosian radius is not available for the entire DECaLS data set. Furthermore, the radii included in the DECaLS catalogue, which are derived by model fitting, were generally found to be much larger than the apparent visible extent of the target source. As a result, the above approach could not be directly replicated for the range of target sources used in the subset. Thus, we needed an alternative method to estimate the appropriate cut-out size for each source.

DESI has objective requirements for optical imaging, one of which is to achieve the magnitude depths of at least 24, 23.4, and 22.5 in the $g$, $r$, and $z$ bands, respectively. These are defined as the `optimal extraction depth' of galaxies near the DESI depth limit. The definition of such a galaxy is an exponential profile with a half-light radius of 0.45 arcsec. An important part of such a profile is the ability to get a good estimate of the number of photometric pixels that make up an image of the galaxy using the equation
\begin{equation}
    N_{\textrm{eff}} = \bigg{[}\bigg{(}4 \pi \sigma^{2}\bigg{)}^{\frac{1}{p}} \; + \; \bigg{(}8.91 r_{\textrm{half}}^{2}\bigg{)}^{\frac{1}{p}}\bigg{]}^{p},
    \label{eq: equation 1}
\end{equation}
\ \\
where $\sigma$ is the standard deviation for a Gaussian fit, p = 1.15 and $r_{\textrm{half}}$ is the half-light radius for an exponential profile fit to the galaxy (Dey et al. \citeyear{Dey_2019}). The equation approximates an exponential fit to a galaxy even though some sources follow a de Vaucouleurs profile, while others have a composite profile. This equation estimates the size of each source, and a cut-out could be extracted as a result. We performed visual tests on a sample of sources to confirm that the estimated cut-out size was sufficient. 

\begin{figure}
    \includegraphics[width=0.95\columnwidth]{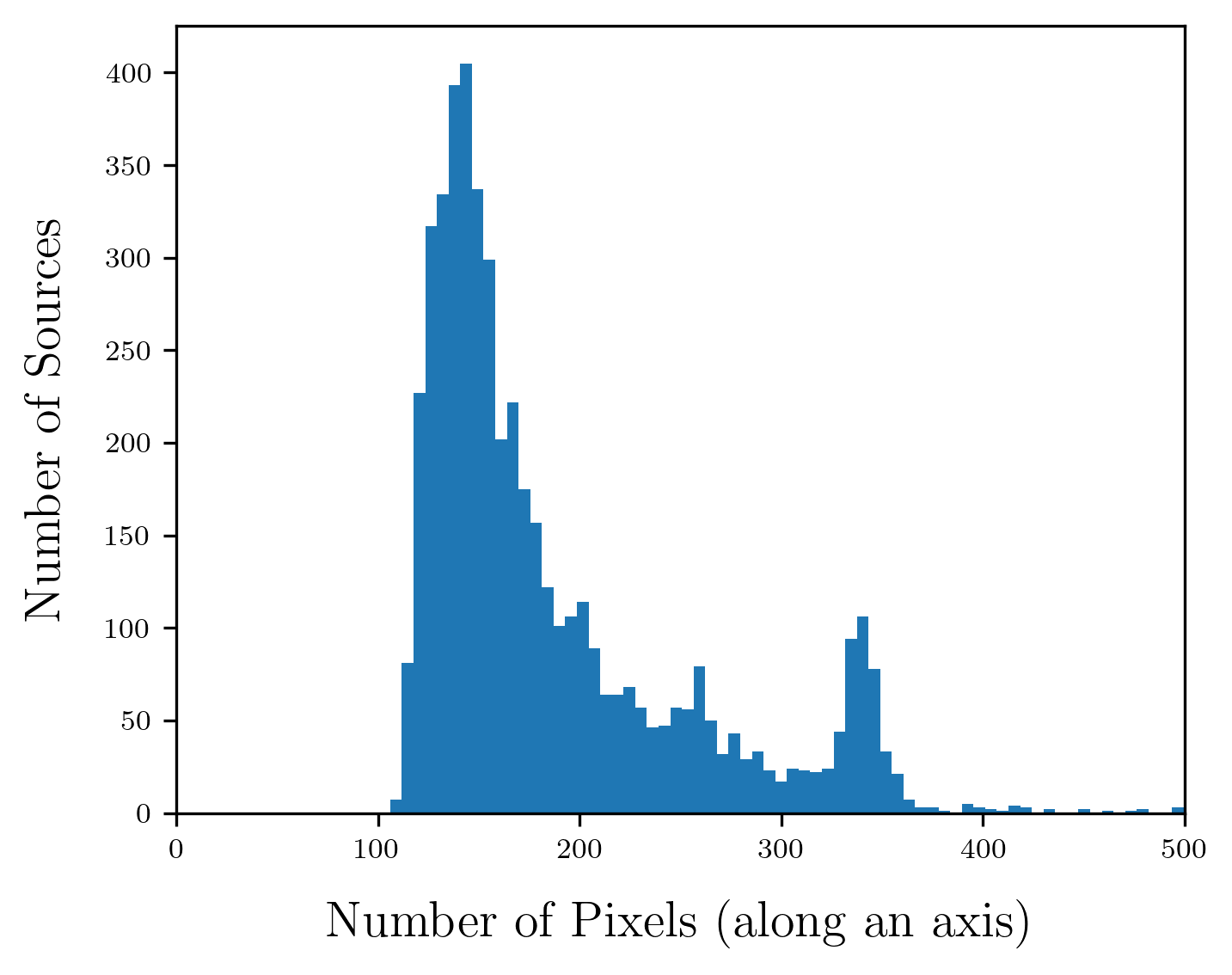}
    \caption{Distribution of the number of sources with varying pixel values as determined by equation (\ref{eq: equation 1}) for a random sample of sources. The number of pixels is given by the square root of $N_{\textrm{eff}}$ in equation (\ref{eq: equation 1}).}\vspace{-20pt}
    \label{fig: Pixel Distribution}
\end{figure}

Fig.  \ref{fig: Pixel Distribution} illustrates the distribution of source numbers based on their respective pixel sizes, as determined by equation equation (\ref{eq: equation 1}), for a random sample of sources. The majority of the sources within this random sample had values between 100 and 200 pixels. To ensure consistency, all source cut-outs used throughout were therefore interpolated to a standardized, yet arbitrary, resolution of 150 $\times$ 150 pixels. This size discrepancy, significantly fewer pixels than the 424 used in Walmsley et al. \citeyearpar{Walmsley_2022}, is attributed to the inclusion of markedly smaller and fainter sources that dominate the population of the subsets investigated. It is worth noting that the secondary peak in Fig \ref{fig: Pixel Distribution}, situated at approximately 350 pixels, corresponds to artefacts present within the random sample used. 

\subsection{Finding an evaluation set}
\label{Data: Finding An Evaluation Set}
A labelled subset of data plays a crucial role in the process of algorithm selection and hyperparameter optimization. Additionally, it serves as a valuable resource for evaluating the performance of anomaly detection techniques. To construct this evaluation subset, we randomly selected 15 000 sources from the complete DECaLS data set and supplemented them with DECaLS cut-outs for 342 lens candidates sourced from Huang et al. \citeyearpar{Huang_2020}. These lenses were intentionally included to ensure the presence of interesting anomalies within the subset.

To ensure consistency and adherence to the criteria described in Section \ref{Data: Selection Cuts}, the same selection cuts were applied to this random set. Table \ref{tab: selection cuts} summarizes the number of sources excluded by each selection cut. We found that a surprisingly high number of the lens candidates did not meet the selection criteria, with only 87 candidates passing all of the applied cuts, implying that the chosen cuts are not optimal for lens inclusion. The majority of failures originated from the Shapedev\_r and Shapeexp\_r cuts, the half-light radii of the de Vaucouleurs and the exponential model respectively, indicating that lenses tend to have small angular diameters and are generally faint. This resulted in sources failing to meet the specified cut criteria, further reinforced by lenses also failing the flux value cut. Some lens candidates failed multiple cuts, and it was observed that three of them were incorrectly labelled as point spread functions (PSFs) in the catalogue. 

\begin{figure}
  \begin{subfigure}[t]{1\linewidth}
    \rotatebox{90}{\hspace{0.65 cm} \Large Label 0}
    \includegraphics[width=0.31\columnwidth]{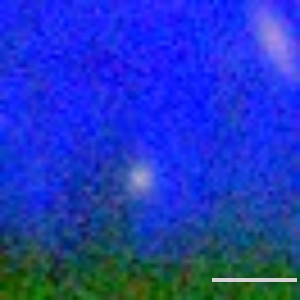}
    \includegraphics[width=0.31\columnwidth]{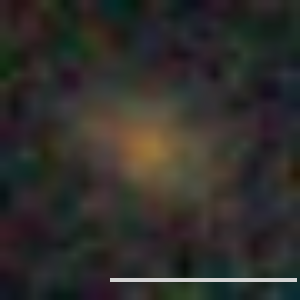}
    \includegraphics[width=0.31\columnwidth]{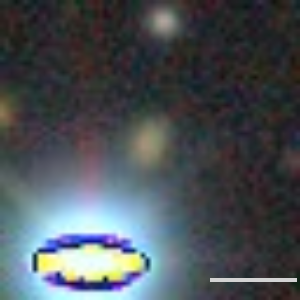}
  \end{subfigure}
  \begin{subfigure}[t]{1\linewidth}
    \rotatebox{90}{\hspace{0.65 cm} \Large Label 1}
    \includegraphics[width=0.31\columnwidth]{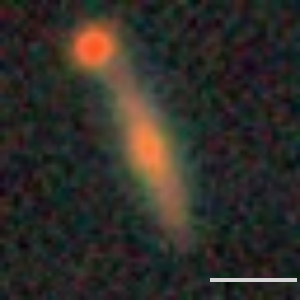}
    \includegraphics[width=0.31\columnwidth]{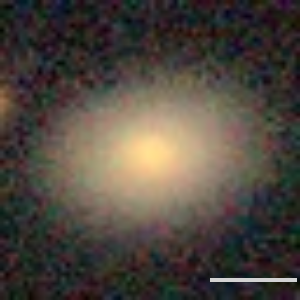}
    \includegraphics[width=0.31\columnwidth]{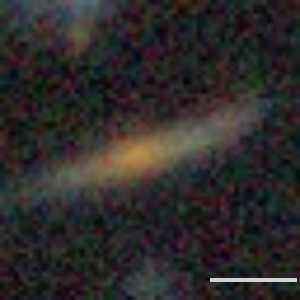}
  \end{subfigure}
  \begin{subfigure}[t]{1\linewidth}
    \rotatebox{90}{\hspace{0.65 cm} \Large Label 2}
    \includegraphics[width=0.31\columnwidth]{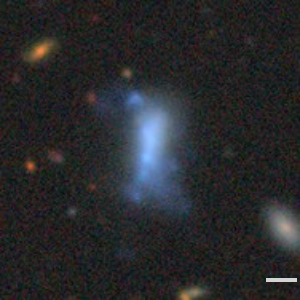}
    \includegraphics[width=0.31\columnwidth]{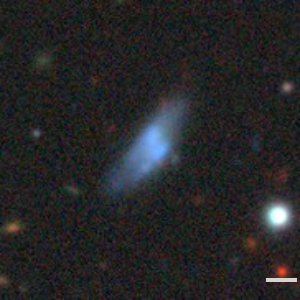}
    \includegraphics[width=0.31\columnwidth]{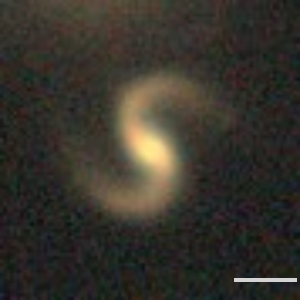}
  \end{subfigure}
  \begin{subfigure}[t]{1\linewidth}
    \rotatebox{90}{\hspace{0.65 cm} \Large Label 3}
    \includegraphics[width=0.31\columnwidth]{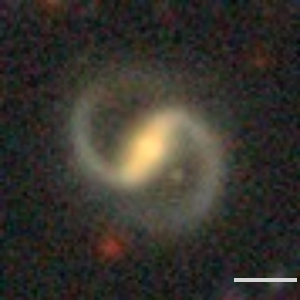}
    \includegraphics[width=0.31\columnwidth]{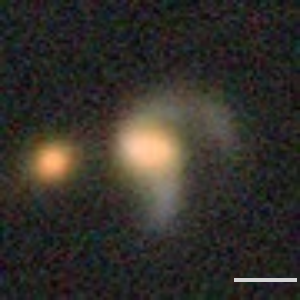}
    \includegraphics[width=0.31\columnwidth]{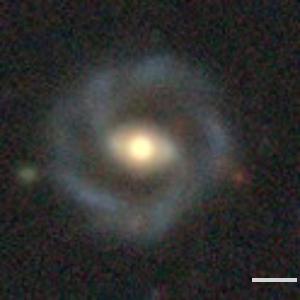}
  \end{subfigure}
  \begin{subfigure}[t]{1\linewidth}
    \rotatebox{90}{\hspace{0.65 cm} \Large Label 4}
    \includegraphics[width=0.31\columnwidth]{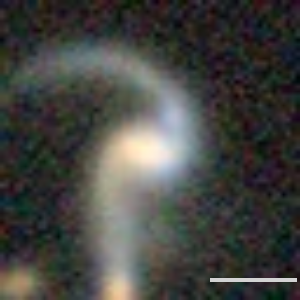}
    \includegraphics[width=0.31\columnwidth]{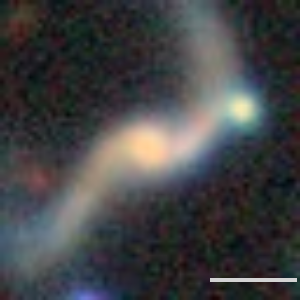}
    \includegraphics[width=0.31\columnwidth]{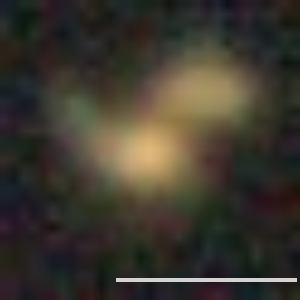} 
  \end{subfigure}
  \begin{subfigure}[t]{1\linewidth}
    \rotatebox{90}{\hspace{0.65 cm} \Large Label 5}
    \includegraphics[width=0.31\columnwidth]{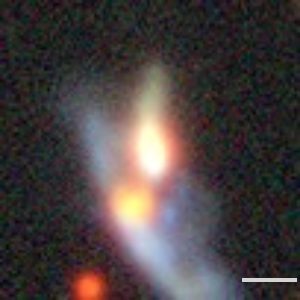}
    \includegraphics[width=0.31\columnwidth]{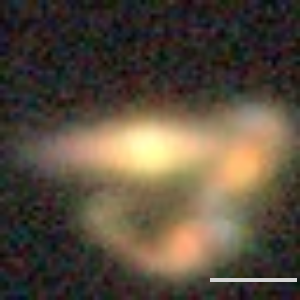}
    \includegraphics[width=0.31\columnwidth]{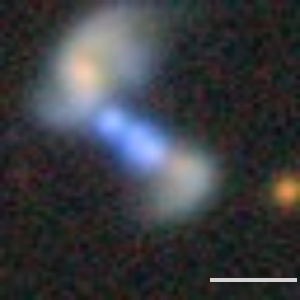}
  \end{subfigure}
    \caption{Examples of sources in the data set with their labels. The first row contains sources such as artefacts, masked sources and low SNR sources. At the bottom are sources that would be considered to be interesting anomalies and contain sources such as galaxy mergers, strong gravitational lenses and other sources that are not readily identifiable.  The angular scale bar within each image represents 5 arcsec.}\vspace{-6pt}
    \label{fig: examples of labels}
\end{figure}

After an extensive search, no generalizable criteria could be found that contain a large sample of lenses while also reducing the size of the entire data set to a manageable level. We thus elected to continue with a sample of 5000 random sources, and the 87 lenses, that passed all selection cuts. This subset of sources was then fully labelled using the labelling scheme described in Section \ref{Methodology} and illustrated in Fig.  \ref{fig: examples of labels}. It is worth noting that even such a small sample is still expected to contain other interesting anomalies, including moderately interesting sources like mergers, in addition to the lenses that have been added to it. This subset formed the evaluation set that was used to assess the performance of the methods used. 

\begin{table}
    \caption{The results of applying different selection criteria to the evaluation set. Each row represents a criterion that was applied. The first column indicates the name of the criterion, the second column shows how many lenses (out of the 342) were excluded by that criterion, and the third column shows how many of the 15 342 sources were excluded by that criterion.}
    \label{tab: selection cuts}
    \begin{tabular*}{\columnwidth}{@{}l@{\hspace*{47pt}}c@{\hspace*{47pt}}c@{}}
        \hline
        \Tstrut
        Selection cut & Lenses & All sources \\
        [2pt]
        \hline 
        \Tstrut
        Type (non-PSF) & 3 & 3 \\ [2pt]
        Allmask & 0 & 0 \\ [2pt]
        Reduced chi-square & 16 & 1840 \\ [2pt]
        SNR & 1 & 1 \\ [2pt]
        Flux values & 19 & 3281 \\ [2pt]
        Shapedev\_r and Shapeexp\_r & 244 & 9513 \\
        [3pt]
        \hline
    \end{tabular*}
\end{table}

\vspace{-10pt}

\section{Methodology}
\label{Methodology}

The methodology is summarized as follows: first, the images are reduced to a representative set of features using a modified CNN followed by a dimensionality reduction procedure (Section \ref{Methodology: Feature Extraction}). These features are then passed to an anomaly detection algorithm, followed by AL to reduce the number of unwanted artefacts and prioritize interesting sources (Section \ref{Methodology: Anomaly Detection}). We evaluate the performance of our methodology using our hand labelled subset of data (Section \ref{Methodology: Evaluation}).

\subsection{Feature extraction}
\label{Methodology: Feature Extraction}
Images are typically high-dimensional and often require dimensionality reduction for computational efficiency. Feature extraction achieves this by transforming images into lower dimensional vectors that retain their essential information. This is done using an image representation function that maps images to vectors while maintaining similarity.

To create the representations of the images that will serve as features in this work, a pre-trained CNN was used, following the approach of Walmsley et al. \citeyearpar{Walmsley_2022}. The CNN model was initially trained on a complex classification task, but its remarkable capability extends to learning relevant features for different tasks beyond its original training purpose.

CNNs have multiple layers, each of which performs a different transformation on the input image. Only the last layer performs classification by generating a probability distribution over the image's possible classes. The rest of the network extracts image information for the classification layer, producing a vector which can be used as features. The CNN in Walmsley et al. \citeyearpar{Walmsley_2022} uses the EfficientNetB0 architecture (Tan \& Le \citeyear{tan2020efficientnet}) and is implemented in \textsc{Zoobot} (Walmsley et al. \citeyear{zoobot}) for various galaxy classification purposes. 

To extract features from the cut-out images, we employed the same model as described by  Walmsley et al. \citeyearpar{10.1093/mnras/stz2816}. However, we used a different pre-processing method to better suit our larger and noisier data set. Since we use the estimated size of the source to extract the cut-out, cropping is not required for our data set. No augmentations of the images were done when passed to the CNN as the network was used for feature extraction only. A standard sigma clipping algorithm from the \textsc{Astropy} python package (The Astropy Collaboration \citeyear{refId0}, \citeyear{Price-Whelan_2018}, \citeyear{TheAstropyCollaboration_2022}) was applied to each image. The algorithm uses an iterative approach to estimate the noise in the image and masks all pixels below the 3$\sigma$ threshold. Finally, the images were greyscaled by averaging the three bands, $g$, $r$, and $z$, into a single band.

The CNN produces a vector that contains 1280 features. This feature vector is still a high-dimensional representation that poses computational challenges. Principal component analysis (PCA) was used to reduce the dimensionality while preserving the majority of the information (Pearson \citeyear{doi:10.1080/14786440109462720}). PCA is a statistical method that transforms a set of correlated variables into a set of uncorrelated variables called principal components, which account for most of the variance in the original data (Shlens \citeyear{Shlens2014}). By setting a variance limit of 95 per cent, the principal components retain 95 per cent of the information in the original features, while reducing the dimensionality from 1280 to 26. After the features were extracted and PCA applied to reduce the dimensionality, an anomaly detection procedure similar to that of Lochner \& Bassett \citeyearpar{Lochner_2021} was applied, using the image representations as features instead of the simple morphological features that were used in that work.

\subsection{Anomaly detection}
\label{Methodology: Anomaly Detection}
\astronomaly{} incorporates two widely used anomaly detection algorithms: isolation forest (iForest) and the local outlier factor (LOF) algorithm (Breunig et al. \citeyear{10.1145/335191.335388}; Liu, Ting \& Zhou \citeyear{10.1109/ICDM.2008.17}) with both algorithms implemented using the \textsc{scikit-learn} package (Pedregosa et al. \citeyear{scikit-learn}). We conducted tests and determined that LOF could not effectively scale to handle the volume of data in the main subset, while iForest proved scalable to such volumes. iForest is a fast algorithm that detects anomalies by employing decision trees. It isolates a data point by randomly selecting a feature and a value within its range, then recursively splits the data into subsections based on this value. This process continues until all data points are isolated, forming a forest of trees. The number of splits required to isolate a point is referred to as the path length, and it serves as a measure of the anomaly score for that point. The underlying idea is that anomalies are more likely to be isolated from the rest of the data, with fewer splits compared to normal points. Therefore, the shorter the path length, the more anomalous the point is considered.

iForest assigns an anomaly score to each source in the data set, determining how it differs from the majority of sources. We normalize this score such that a higher value represents how anomalous the object is and thus determine a ranked order of the entire subset from most to least anomalous. AL can then be applied, where the graphical interface of \astronomaly{} allows the users to manually rate the objects on a scale of 0 to 5 according to how interesting they are. 

AL is an approach that allows algorithms to select the most informative examples to label in order to improve model performance. The classification of anomalies is often subjective and depends on the user's judgement to identify objects of particular interest. By focusing on the relevant samples that provide the most information, AL aims to reduce the amount of labelled data needed. This is useful in situations where labelling data is expensive and time-consuming and requires human expertise, or where no labels exist and only a subset of the data is sufficient to achieve good results. \astronomaly{} uses these user-provided labels to update the anomaly scores for the entire data set, leading to a revised ranking order. 

Two different AL approaches were evaluated in this work. The first approach is the novel AL algorithm introduced by Lochner \& Bassett \citeyearpar{Lochner_2021}, which is integrated into \astronomaly{} and referred to in this study as the neighbour score (NS) algorithm. The second is the one used in Walmsley et al. \citeyearpar{Walmsley_2022}, a GP (Rasmussen \& Williams \citeyear{rasmussen2006gaussian}) that can model smooth distributions, hereon referred to as the direct regression (DR) algorithm.

The primary objective of the NS algorithm is to adjust anomaly scores based on user-provided labels. It uses training data consisting of a small number of human-provided labels and employs a random forest regression algorithm (Liaw \& Wiener \citeyear{Randon_Forest}) to predict user scores for all data instances based on these labels. By calculating each instance's distance from its nearest labelled neighbour, the method effectively identifies areas of the feature space where the algorithm displays uncertainty. In these regions, it returns scores that are close to the original anomaly scores. Conversely, in regions with ample training data, the predicted user scores modulate the anomaly scores. The algorithm employs a scoring function that combines nearest neighbour distances, predicted user scores, and original anomaly scores to compute the final anomaly scores for each instance.

In contrast, the DR approach skips entirely the truly unsupervised anomaly detection step. Instead, it attempts to use AL to directly model the user's `interestingness' score. The approach of Walmsley et al. \citeyearpar{Walmsley_2021} is to select a small set of random examples for labelling and then iteratively query the user (sometimes called the oracle in machine learning literature) with a new set of examples explicitly selected, using an `acquisition' function, in order to improve the regression algorithm. While any regression algorithm can be used, it must be able to produce uncertainty information in order to compute the acquisition function. GPs are ideally suited to this task as they model uncertain or noisy functions. They assign probabilities to potential data-fitting functions and update these probabilities as data points are labelled, transitioning from a prior to a posterior distribution. GPs rely on a kernel to shape function behaviour and smoothness, with hyperparameters adjusted based on data likelihood to maximize the fit to observed data.

The GP incorporated in \astronomaly{} uses a combination of the Mat\'ern kernel and the WhiteKernel to model the relationships between data points and account for noise in the predictions. Both implementations are implemented in Python through the \textsc{scikit-learn} package (Pedregosa et al. \citeyear{scikit-learn}).

\vspace{-10pt}

\subsection{Evaluation}
\label{Methodology: Evaluation}
To evaluate the performance of the two AL approaches, the fully labelled evaluation set described in Section \ref{Data: Finding An Evaluation Set} was used as a benchmark, which allowed the recall of the algorithms to be determined. iForest was applied to compute the anomaly score for each data point in the data set. In the NS algorithm, the process involved selecting the top 100 images with the highest anomaly scores. These images were then labelled through the \astronomaly{} interface to establish a new ranking order based on user input. From this new ranking, the top 100 sources were identified and labelled, excluding those already labelled in previous iterations. This iterative process continued until a total of 500 sources were labelled and used for training. For the DR algorithm, the first 100 sources were manually labelled based on iForest scores, and the sources were then sorted by acquisition scores. The retraining procedure was repeated iteratively until 500 sources had been labelled and trained upon. Throughout this process, the accuracy of each labelling iteration was assessed to evaluate the performance and effectiveness of the algorithms.

Fig.  \ref{fig: examples of labels} illustrates typical sources with different labels (ranging from 0 to 5) assigned as scores during the AL process. Score 0 includes artefacts, masked sources and low SNR sources. Scores 1 and 2 contain typical elliptical and spiral galaxies respectively. Score 3 showcases more interesting spiral galaxies with unusual structures. Score 4 is reserved for sources with highly unusual morphologies, while score 5 contains clearly interacting sources, lenses, and other unidentified sources considered anomalous. It is important to note that labelling is subjective and that this served as a guideline only. It does not comprehensively capture all variations within the data set, nor should it be viewed as a rigid classification system.

\vspace{-10pt}

\begin{figure}
    \includegraphics[width=\columnwidth]{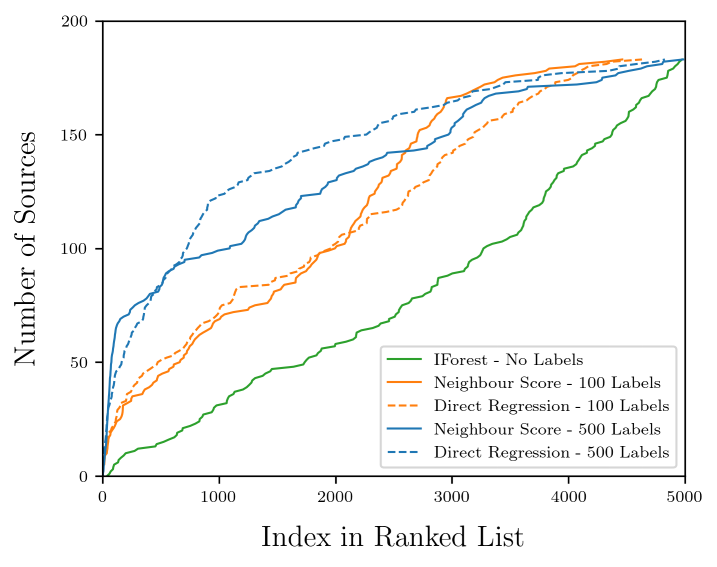}
    \caption{Performance of the NS and DR active learrning algorithms on the evaluation set. The algorithms were applied in iterations of 100 labels, with the results from 100 and 500 labels illustrated. Both algorithms have comparable performance.}\vspace{-10pt}
    \label{fig: AL vs GP for all anomalies}
\end{figure}

\vspace{-10pt}

\section{Results}
\label{Results}
To evaluate the performance of anomaly detection methods, particularly in scenarios where the data set contains an unknown number of interesting anomalies, we employ two types of plots: a recall plot and a Uniform Manifold Approximation and Projection (UMAP) plot (McInnes, Healy \& Melville \citeyear{mcinnes2020umap}).

Recall plots, as depicted in Fig.  \ref{fig: AL vs GP for all anomalies} and Fig.  \ref{fig: Normalised plot}, serve to illustrate how the anomaly detection algorithm ranks sources from most to least anomalous based on their anomaly scores. In this plot, the x--axis represents the position of the sources in the ranked list, while the y--axis represents the recall for the specific class of interest. Essentially, the y--value increases when a source belongs to the class of interest. A steeper slope at the beginning of the plot indicates better performance, as it means that a user has to search through fewer sources to find interesting anomalies. This ranking of sources is crucial because, in the \astronomaly{} interface, the sources on the left would be displayed first for labelling when using the score-based ranking. Other ordering methods, such as random ordering, are also available in the interface. 

To visualize the high-dimensional features in a two-dimensional space, the UMAP technique was used in this work (McInnes, Healy \& Melville \citeyear{mcinnes2020umap}). UMAP is a non-linear dimensionality reduction technique that retains the topological structure of the data. This means it can capture both global and local relationships among data points. In a UMAP plot, clusters are groups of points that are close together in feature space, suggesting that the data points within the clusters share similarities or patterns that differentiate them from other data points. UMAP can also highlight outliers, which are data points that significantly differ from the majority of the data. Outliers may appear as isolated points far from any clusters or as points located on the edges of clusters but not tightly grouped with them. By embedding high-dimensional features into a two-dimensional space, UMAP visualizations, like the one shown in Fig.  \ref{fig: UMAP of evaluation set}, help provide valuable insights into the underlying data structure and the performance of the anomaly detection methods.

\subsection{Evaluation set}
\label{Results: Evaluation Set}
The comparison of the NS, DR, and iForest-only anomaly detection methods is illustrated in Fig.  \ref{fig: AL vs GP for all anomalies}. The plot shows how many of the known anomalies in the evaluation set were detected as a function of the sample size. The evaluation set consists of a total of 184 anomalies including the 87 lenses injected from a known catalogue.
The performance of iForest on the evaluation set yielded somewhat surprising results, as it detected only 15 anomalies among the top 500 sources. However, when the DR approach was applied and trained with 100 labels, it detected 51 anomalies, while the NS method found 45. Both AL methods demonstrated a significant improvement in anomaly detection compared to using iForest alone. Further increasing the number of labels to 500  resulted in even better anomaly detection performances, with both AL methods detecting 84 anomalies within the top 500 sources.

\begin{figure}
    \includegraphics[width=\columnwidth]{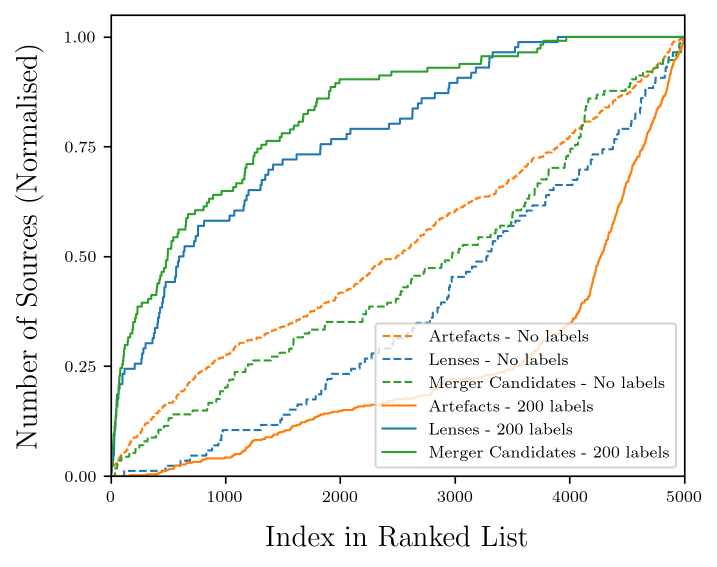}
    \caption{Recall of three types of anomaly in the evaluation set. The dashed lines in the plot represent the initial results from iForest, while the solid lines represent the results obtained from the NS algorithm trained with 200 labels. The plot has been normalized to emphasize the performance gains of AL. AL improves the detection rates of interesting sources while decreasing the impact of unwanted artefacts.}\vspace{-6pt}
    \label{fig: Normalised plot}
\end{figure}

To gain a deeper understanding of why iForest struggled to detect a significant number of anomalies, a more detailed analysis was conducted on the sources and their associated scores. Fig.  \ref{fig: Normalised plot} presents a normalized plot of the sources in the evaluation set, categorized into three distinct classes: artefacts, galaxy merger candidates, and known lenses, as these were the sources of particular interest. To directly compare classes with very different numbers of objects, we normalize the recall in each class to a scale of 0 to 1. This plot reveals the reason for iForest's poor performance: sources with a high anomaly score are predominantly artefacts. While technically anomalies, these sources may not be particularly interesting from a scientific perspective. 

Fig.  \ref{fig: Normalised plot} provides additional evidence of the enhanced detection rates achieved through the application of the NS algorithm. The galaxy mergers and gravitational lensing candidates have been shifted higher in the order (to the left and upwards), while the artefacts have been moved lower (down and to the right). This outcome is quite important, as it demonstrates that AL can rapidly remove artefacts that may have been missed by automated pipelines, allowing a scientist to focus their attention on interesting anomalies.

\begin{figure*}
\vspace{-12pt}
\begin{minipage}{\textwidth} 
  \twocolumn[{%
    \centering
    \includegraphics[width=0.98\textwidth]{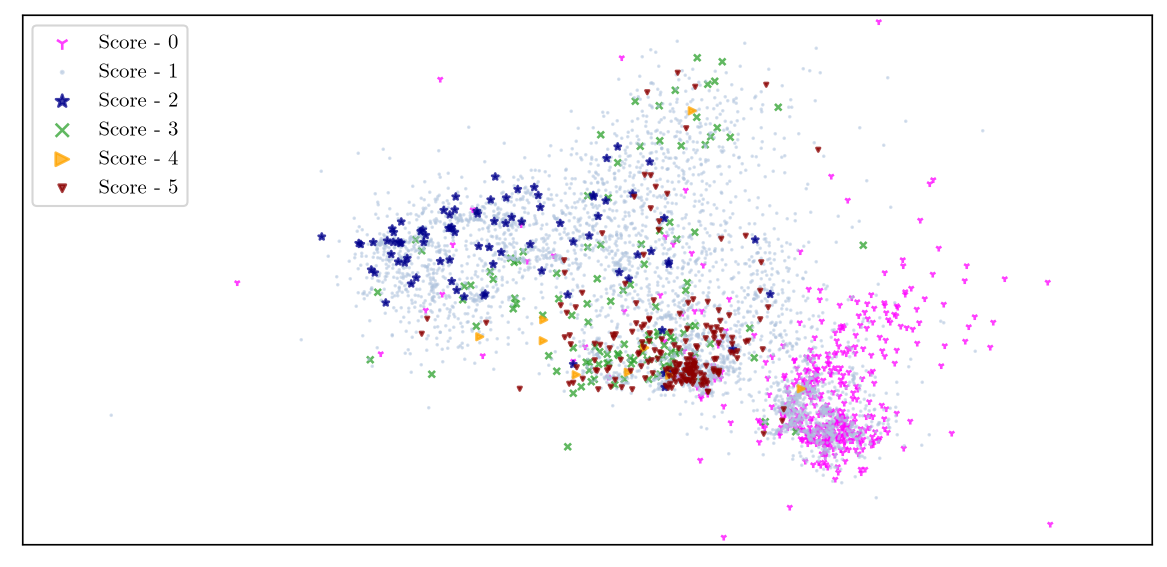}
    \vspace{-5pt}
    \caption{UMAP plot of the evaluation set. The different scores/labels show that subclusters are formed within the feature space, but that they are surrounded by the more common, uninteresting sources. It is interesting to note that the artefacts present, represented in pink, form a relatively distinct cluster. The anomalies show a similar pattern, with a very dense cluster formed.}\vspace{-5pt}
    \label{fig: UMAP of evaluation set}
    \strut
    }]
\end{minipage}
\end{figure*}

Fig.  \ref{fig: UMAP of evaluation set} presents a UMAP plot for the evaluation set, where all sources were labelled and could therefore be classified into different `classes' based on their scores. The most prevalent class, represented by a Score of 1, typically includes `normal' galaxies with no specific interest. All other classes formed distinct sub-clusters within the UMAP plot, showing minimal overlap with each other. Of special interest are the classes associated with artefacts (Score 0) and the most interesting anomalies (Score 5). These two classes display dense clusters, and the artefacts are notably well-separated from all other sources. First, this demonstrates the capability of image representations to effectively extract features that distinguish classes from one another. Secondly, this offers valuable insights into the challenges encountered when using iForest as a standalone anomaly detection method. The interesting anomalies are located on the edge of the overall structure of the plot, but the artefacts are much more likely to be detected as anomalies by iForest given the position and distance of their sub-cluster. These sub-clusters highlight the importance of AL algorithms, which identify these regions of interest in the feature space and are crucial for eliminating artefacts. 

The feature space in Fig.  \ref{fig: UMAP of evaluation set} is markedly different from that of Walmsley et al. \citeyearpar{Walmsley_2022}, where most of the interesting anomalies were deep in the centre of the plot. This is likely due to the fact that the Galaxy Zoo data represent a specific subset of DECaLS with very different properties. The Galaxy Zoo data set is well curated and subject to more restrictive magnitude cuts, among other criteria, resulting in the selection of galaxies that are well-resolved in general. This results in a very different structure in the feature space which explains why iForest, combined with the NS algorithm, worked well here but failed to find interesting anomalies in the Galaxy Zoo data with the CNN-based features.

All of these classes are interspersed among the common sources, those with a Score of 1, indicating that it is quite diverse and contains features common with all of the other classes. The class with a Score of 3 is closest to the interesting anomalies, which is likely due to the similarities between spiral galaxies with moderately interesting morphologies (which constitute most of the Score 3 class) and the galaxy merger candidates in the Score 5 class. The Score 4 class contains too few sources to draw clear observations about it. 

\begin{figure*}
\vspace{-6pt}
\begin{minipage}{\linewidth}
    \centering
    \includegraphics[width=0.48\textwidth]{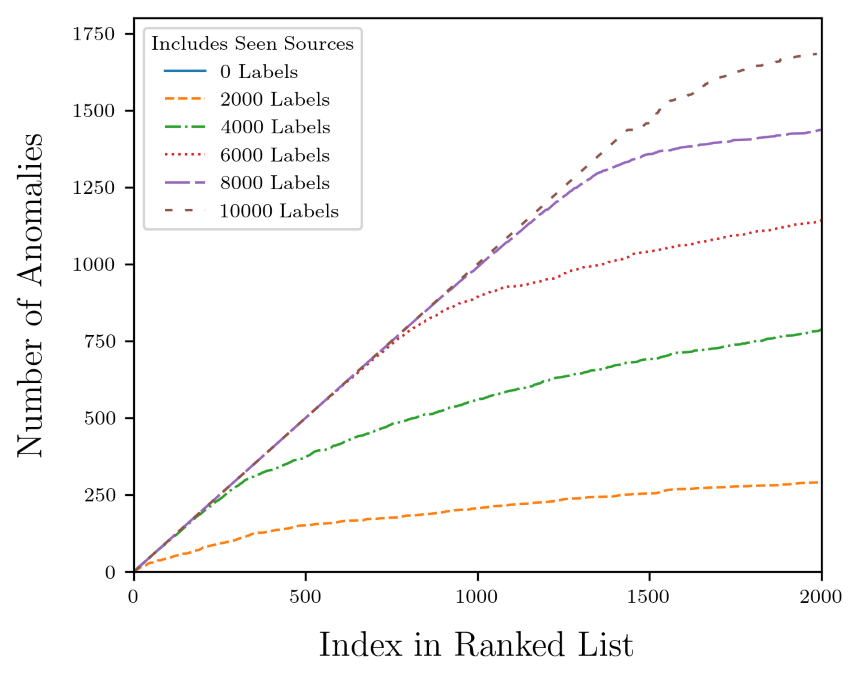}
    \includegraphics[width=0.48\textwidth]{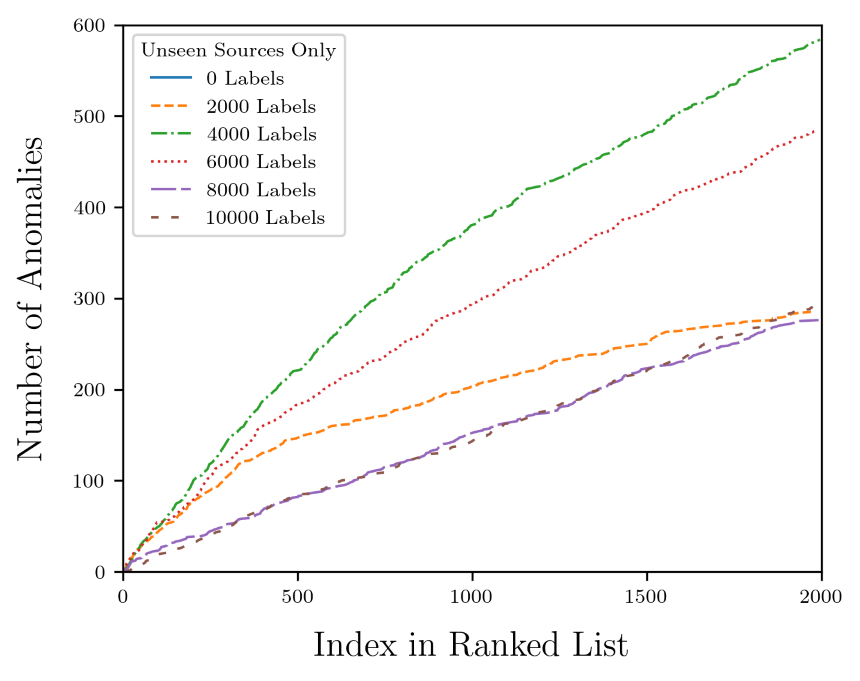}
    \vspace{-4pt}
    \caption{Recall as a function of rank for increasing numbers of labels. The plot on the left shows the number of anomalies including those that have been previously labelled, whereas the plot on the right excludes the labelled anomalies, showing the unlabelled sources the algorithm determines as being the most likely to be interesting. It is clear that the number of ``new'' anomalies found in the top 2000 increases as more labels are used for training. However, there is a point of diminishing returns as the number of ``new'' anomalies decreases rapidly when more than 4000 labels are used. It should also be noted that the line for 0 labels is not visible on either plot since there was only one source detected.}
    \label{fig: Anomalies in top 2000: labelled}    
\end{minipage}
\end{figure*}

\vspace{-10pt}

\subsection{Application on a large scale}
\label{Results: Application on a large scale}
The NS algorithm was used in the application on the main subset, which consists of 3 884 404 unlabelled sources. The volume of this subset forms the main challenge of this work as \astronomaly{} has not been applied on such a scale before. The same approach as in the evaluation set was followed, excluding the use of the DR algorithm due to its computational demands and limited discernible benefits. See Appendix \ref{Appendix: Computational Resources} for more details. Each AL iteration of the NS algorithm was done with 2000 labels until 10 000 sources were labelled. The top 2000 sources were investigated and fully labelled in each iteration. This entire process only took the lead author several hours using \astronomaly{}'s interactive interface.

The results are presented in Fig.  \ref{fig: Anomalies in top 2000: labelled}, showing the anomalies and their ranks among the top 2000 sources. The left panel includes labelled sources and the right panel shows only the new or unseen anomalies, demonstrating the power of the algorithm to detect new anomalies. The line for 0 labels, corresponding to iForest only, is not visible because it detected only one interesting anomaly in the top 2000 sources. However, this does not mean that iForest failed to detect anomalies. In fact, most of the 2000 (1763) sources initially detected by iForest were artefacts or masked sources; obvious anomalies in the data set, but not of interest. 
The plots demonstrate the necessity of AL, as more labels lead to a higher number of interesting sources within the top 2000. However, as the panel on the right in Fig.  \ref{fig: Anomalies in top 2000: labelled} shows, the number of new anomalies seen in the top 2000 drops sharply when more than 4000 labels are used. This suggests a point of diminishing returns, where adding more labels would not necessarily lead to more anomalies being found when looking at the same number of sources. In such an instance, discovering additional anomalies would require investigating a larger number of sources rather than increasing the number of labels.

\subsection{Follow-up investigations}
\label{Results: Follow-up investigations}
The 10 000 labels from the main subset were scored between 0 and 5 using the labelling scheme described in Section \ref{Methodology} and illustrated in Fig.  \ref{fig: examples of labels}. The majority of the sources (4861) received a score of 0, indicating that most were artefacts or masked sources. The next largest group (2408) that received a score of 1 were the common galaxy types, followed by 1648 sources given a score of 5, indicating  a large number of interesting anomalies. The other labels were distributed among the scores of 2 (519), 3 (288), and 4 (276), which represent galaxies with slightly disturbed morphology. 

The 1648 anomalies were analysed and 18 unclassified sources (Fig.  \ref{fig: Anomalies unusual}), eight gravitational lens candidates (Fig.  \ref{fig: Anomalies lenses}) and 1609 galaxy merger candidates (Fig.  \ref{fig: Anomalies mergers}) were detected. Out of the 1648 sources labelled to be anomalies, further investigations determined that 13 of these sources were either artefacts, other uninteresting sources mislabelled as lens candidates or sources that form part of another detected source and which could be considered to be a duplicate detection. In an attempt to identify these anomalies, they were cross-matched with the Simbad data base (Wenger et al. \citeyear{Wenger_2000}) with different cross-matching distances and using all of the available data sets on Simbad. A total of 1209 matches at 30 arcsec and 792 matches at 10 arcsec were obtained, generally corresponding to already labelled galaxies. However, there was no definitive data set to match against that could confirm the nature of the anomalies detected and a manual follow-up would be necessary to confirm their nature, as even a 10 arcsec difference can lead to potential mismatches. 

\vspace{-10pt}

\subsubsection{Sources with highly unusual morphology}
\label{Results: Unusual Sources}
\begin{figure}
\begin{minipage}{\columnwidth} 
    \includegraphics[width=0.325\columnwidth]{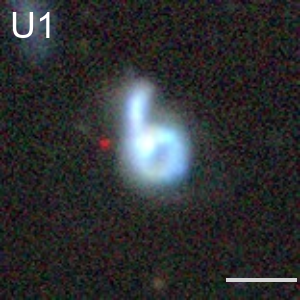} \hfill 
    \includegraphics[width=0.325\columnwidth]{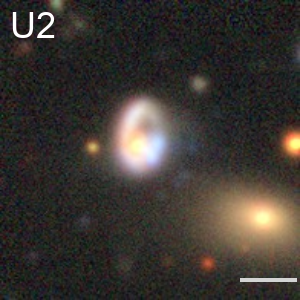} \hfill
    \includegraphics[width=0.325\columnwidth]{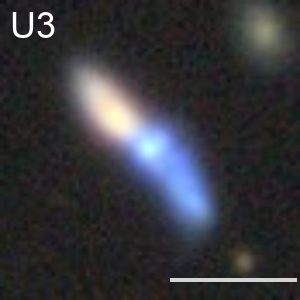} \hfill \\
    \includegraphics[width=0.325\columnwidth]{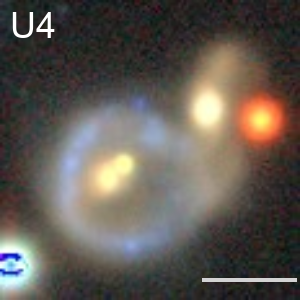}   \hfill 
    \includegraphics[width=0.325\columnwidth]{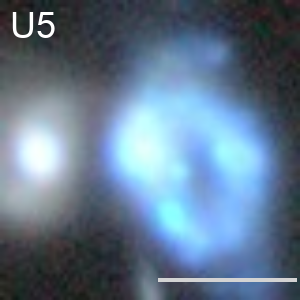} \hfill
    \includegraphics[width=0.325\columnwidth]{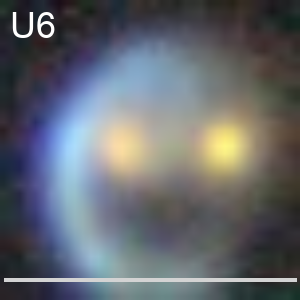}  \hfill \\
    \includegraphics[width=0.325\columnwidth]{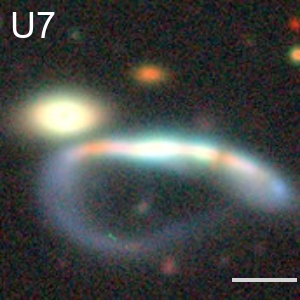} \hfill
    \includegraphics[width=0.325\columnwidth]{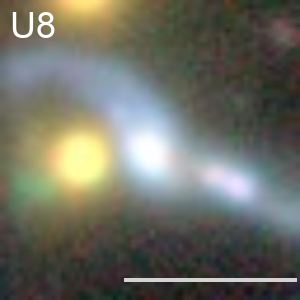} \hfill 
    \includegraphics[width=0.325\columnwidth]{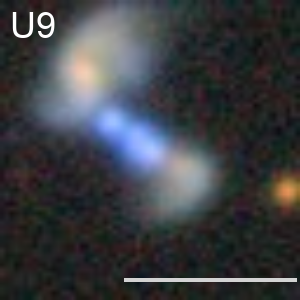} \hfill \\
    \includegraphics[width=0.325\columnwidth]{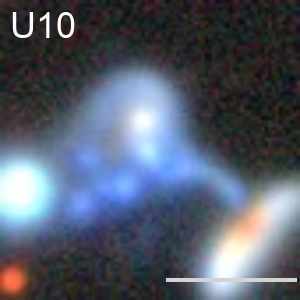}  \hfill 
    \includegraphics[width=0.325\columnwidth]{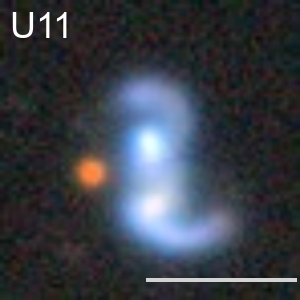} \hfill 
    \includegraphics[width=0.325\columnwidth]{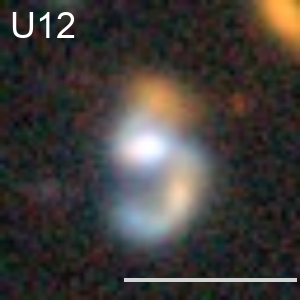} \hfill \\
    \includegraphics[width=0.325\columnwidth]{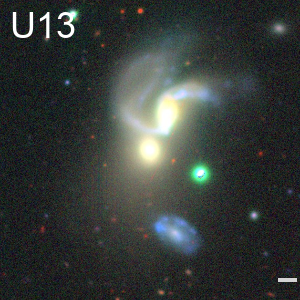} \hfill 
    \includegraphics[width=0.325\columnwidth]{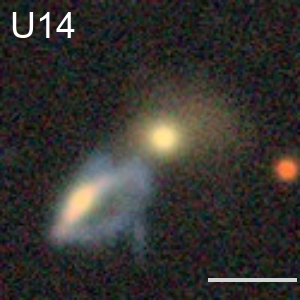}  \hfill 
    \includegraphics[width=0.325\columnwidth]{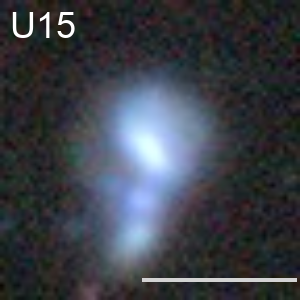} \hfill \\
    \includegraphics[width=0.325\columnwidth]{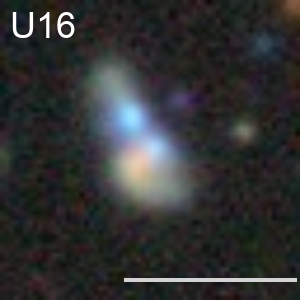} \hfill  
    \includegraphics[width=0.325\columnwidth]{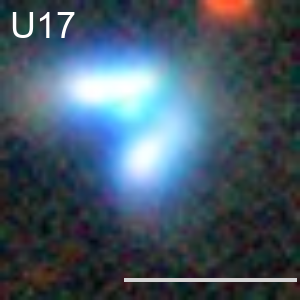} \hfill 
    \includegraphics[width=0.325\columnwidth]{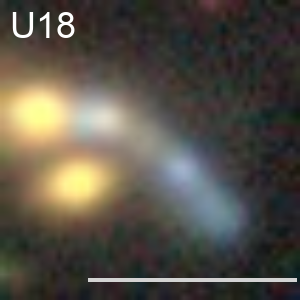}  \hfill \\ 
    \vspace{-2pt}
    \caption{The anomalies shown here are difficult to identify with a quick visual inspection. They need further investigations to determine their nature and origin. Table \ref{tab: Follow-up investigations} lists more information on these anomalies. The angular scale bar within each image represents 10 arcsec.} \vspace{-6pt}
     \label{fig: Anomalies unusual}
\end{minipage}
\end{figure}

We searched for matches in Simbad (Wenger et al. \citeyear{Wenger_2000}) for the 18 unidentified sources. Table \ref{tab: Follow-up investigations} summaries the source locations and any known identifiers with a link to the Simbad entry. Almost half of the sources have no matches or identifications in any other data sets. The sources can be seen in Fig.  \ref{fig: Anomalies unusual} along with their labels.

Here we summarize our initial investigations of the sources, based on information found in Simbad and the Data Aggregation Service\footnote{\url{https://das.datacentral.org.au/das}}. 

\begin{table}
 \caption{Information about the 18 initially unidentified anomalous sources detected in the main set. The second and third columns show the right ascension and declination in degrees, respectively. The fourth column lists any known source names of the target from other surveys or catalogues if they were matched. Blank entries are unmatched sources.}
\begin{tabular*}{\columnwidth}{@{\hspace*{0pt}}l@{\hspace*{18pt}}c@{\hspace*{18pt}}c@{\hspace*{18pt}}c}
    \hline 
    \Tstrut Entry & RA & \textrm{Declination} & Identifiers \\
          & [deg] & \textrm{[deg]} &  \\
    [4pt] 
    \hline
    \Tstrut
    U1 & 209.7286 & $29.5764$  & \href{http://simbad.u-strasbg.fr/simbad/sim-coo?Coord=209.7290d29.5764d&CooFrame=ICRS&CooEpoch=2000&CooEqui=2000&Radius=2&Radius.unit=arcmin&submit=submit+query&CoordList=}{2MASS J13585490+2934356}  \\
    U2 & 342.9047 & $17.8460$  & \href{http://simbad.cds.unistra.fr/simbad/sim-id?Ident=%401459688&Name=2MASX%20J22513724%2b1750457&submit=submit}{2MASX J22513724+1750457} \\ 
    U3 & 60.7336  & $-15.2434$ & \href{http://simbad.u-strasbg.fr/simbad/sim-coo?&Coord=60.733628636793d-15.243399852026d&CooFrame=FK5&Radius=2&Radius.unit=arcmin}{LEDA 913772} \\ [2pt]
    U4 & 42.2626  & $3.2043$   &  \\ 
    U5 & 27.0134  & $-21.6656$ & \href{http://simbad.cds.unistra.fr/simbad/sim-id?Ident=%401279898&Name=ESO%20543-16&submit=submit}{ESO 543-16}  \\ 
    U6 & 22.1821  & $-2.3705$  &  \\ 
    U7 & 69.8371  & $-50.5307$ & \href{http://simbad.cds.unistra.fr/simbad/sim-id?Ident=%402998689&Name=ESO%20202-45&submit=submit}{ESO 202-45}  \\ 
    U8 & 31.6507  & $-28.0090$ & \href{http://simbad.cds.unistra.fr/simbad/sim-id?Ident=%404726613&Name=2dFGRS%20TGS226Z042&submit=submit}{2dFGRS TGS226Z042} \\ 
    U9 & 50.7173  & $-11.8873$ &  \\ 
    U10 & 310.1343 & $1.8177$   &  \\ 
    U11 & 63.0540  & $-24.9667$ &  \\ 
    U12 & 36.2380  & $-27.2902$ & \href{http://simbad.cds.unistra.fr/simbad/sim-id?Ident=%404702435&Name=2dFGRS%20TGS230Z092&submit=submit}{2dFGRS TGS230Z092} \\   
    U13 & 210.4254 & $33.8215$  & \href{http://simbad.cds.unistra.fr/simbad/sim-id?Ident=%405133343&Name=NVSS%20J140141%2b334937&submit=submit}{NVSS J140141+334937} \\ 
    U14 & 212.8158 & $0.1169$   & \href{http://simbad.cds.unistra.fr/simbad/sim-id?Ident=%402238031&Name=SDSS%20J141116.31%2b000654.9&submit=submit}{SDSS J141116.31+000654.9} \\
    U15 & 46.9290  & $-14.1055$ & \href{http://simbad.u-strasbg.fr/simbad/sim-coo?&Coord=46.9290218d-14.1056227d&CooFrame=FK5&Radius=2&Radius.unit=arcmin}{LEDA 928927} \\
    U16 & 55.5332  & $-19.8353$ &  \\
    U17 & 106.5698 & $68.5793$  &  \\ 
    U18 & 32.2221  & $-0.9785$  & \href{https://simbad.cds.unistra.fr/simbad/sim-id?Ident=%4013689881&Name=SDSS%20J020853.45-005841.1&submit=submit}{SDSS J020853.45-005841.1}  \\
    [2pt]
    \hline
    \end{tabular*}
    \label{tab: Follow-up investigations}
\end{table}

\begin{enumerate}[label=\indent(\roman*), align=left, itemindent=14pt, labelwidth=0.0em]
    \item U1 -- This appears to be a ring-shaped starburst galaxy (Toba et al. \citeyear{2014ApJ...788...45T}) at a spectroscopic redshift of 0.077 (Ahumada et al. \citeyear{SDSS2020}). The northern extension may be a tidal tail or interacting galaxy, but may also be coincident as it has an imprecise photometric redshift of 0.11.
    \item U2 -- This puzzling object has detectable emission in radio \nvss{}, infrared \twomass{}, and ultraviolet \galex{}. Two competing spectroscopic redshifts are available, 0.071 \glade{} and 0.2 \citep[Milliquas,][]{Flesch2023}. The latter seems to correspond to the bright red source which could be a coincident quasar (given the selection criteria of Milliquas). A detailed analysis would be required to determine which of these sources are associated and if the ring is a lens or an unusually red ring galaxy with a bright patch of star formation.
    \item U3 -- This source, apparently half red and half blue, has the same visual appearance in PanSTARRS \citep{Chambers2016} indicating its appearance is not due to a processing error. Unfortunately, only photometric redshifts from DECaLS are available with an implied redshift of 0.1120 for the red source and 0.1050 for the bright blue bulge in the centre. The uncertainties are too large to determine if these are coincident sources or not. The entire source has associated ultraviolet emission from GALEX and the red part is clearly detected in 2MASS. We have no immediate explanation for the dual-colour nature of this source. It could be two coincident galaxies with a remarkable chance alignment, interacting galaxies of very different colours and nature or something else entirely.
    \item U4 -- This source has a star-forming ring and two apparent cores that may be merging. A photometric redshift of 0.0775 \glade{} is available for one of the cores. The nearby galaxy, \href{http://simbad.u-strasbg.fr/simbad/sim-coo?&Coord=42.263805d3.203656d&CooFrame=FK5&Radius=2&Radius.unit=arcmin}{LEDA 213095}, has a photo-z of 0.0734 \glade{} meaning it could potentially be interacting and triggering the star formation.
    \item U5 -- A highly disturbed, star-forming ring galaxy. Not surprisingly, this object is bright in radio \nvss{} and ultraviolet \galex{}. Only the bright ring has spectroscopic information, with a redshift of 0.0459 \sixdf{}. The neighbouring source has a negative photo-z from DECaLS. It is thus not clear if these galaxies are interacting or coincident. At the bottom of the cut-out, a faint third galaxy can be seen for which no redshift information is available.
    \item U6 -- This object appears to be a lens but has the unusual feature of two foreground objects, neither of which is in the centre. A competing explanation is that the pair of galaxies is disturbing a third galaxy, completely disrupting its morphology. The left and right foreground galaxies have photometric redshifts of $0.3117\pm0.1321$ and $0.1224\pm0.0377$ respectively which suggests that one of them may be coincident. However, given the complexity of the system and the large uncertainty of the redshift of the left source, we cannot be certain they are unrelated. The entire system is visible in ultraviolet \galex{} and the two foreground sources are faintly detected in infrared \twomass{}.
    \item U7 -- Tidal tail due to interaction with a neighbouring galaxy. Both galaxies have a redshift of 0.0371 \glade{}, although a redshift error is not listed so they may still be coincident sources.
    \item U8 -- It is difficult to determine if the two red galaxies are associated with the oddly shaped blue galaxy. The redshift estimates for the parts of this system vary significantly, although the DECaLS photometric redshifts place the red galaxies at a similar redshift (0.22) to the spectroscopic redshift of the blue galaxy \sixdf{}. However, the group-finding algorithm of \citet{Eke2004} places these galaxies in a group at redshift 0.11. It seems likely that it is the group interaction that has warped the morphology of the blue galaxy triggering star formation, but careful data analysis is needed to confirm this.
    \item U9 -- The two galaxies seen here have a similar photometric redshift in the DECaLS catalogue of 0.17, but the blue region in the middle registers a very different redshift of 0.34. It is possible that these galaxies are simply coincident or they may in fact be interacting and creating a star-forming region between them which is not well estimated by photo-z algorithms.
    \item U10 -- With very different redshift estimates for each part of this system, these sources could very well be coincident. However, it seems quite a dramatic alignment of sources so some amount of interaction and star formation may be more plausible. Additional spectroscopic observations would be needed to determine the nature of this interesting group.
    \item U11 -- This unusual system also has disagreeing photometric redshifts so it may be a chance alignment, although it visually appears to be a merging system.
    \item U12 -- This unusual group of galaxies has been detected with the group-finding algorithm of \citet{Eke2004} at redshift 0.2137.
    \item U13 -- This system was identified as a possible compact group in \citet{Zheng2020} at redshift 0.02640 with possible interactions. The main galaxy is also the host of the supernova SN 2012T \citep[Asiago Supernova Catalogue,][]{Barbon1999}.
    \item U14 -- Known group of galaxies at redshift 0.1615 \citep{Eke2004}.
    \item U15 -- With three different photometric redshift estimates in this group, these are likely to be coincident galaxies although the usual caveats about photo-z uncertainties apply.
    \item U16 -- All three photometric redshifts from DECaLS for this system are very different so again, they are either coincident galaxies or have poorly estimated redshifts.
    \item U17 -- Likely chance alignment due to differing photometric redshifts.
    \item U18 -- Disagreeing spectroscopic redshift estimates, \href{https://simbad.cds.unistra.fr/simbad/sim-id?Ident=%403788069&Name=%5bPVK2003%5d%20032.22307%20-00.97894&submit=submit}{0.19623} and \href{https://simbad.cds.unistra.fr/simbad/sim-id?Ident=%4013689881&Name=SDSS%20J020853.45-005841.1&submit=submit}{0.17436}, suggest these may be colinear galaxies.
\end{enumerate}

\vspace{-10pt}

\subsubsection{Gravitational lens candidates}
\label{Results: Gravitational Lens Candidates}
\begin{figure}
\begin{minipage}{\columnwidth} 
\centering
    \includegraphics[width=0.48\columnwidth]{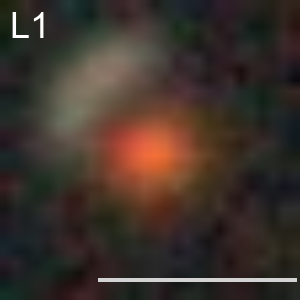}   
    \includegraphics[width=0.48\columnwidth]{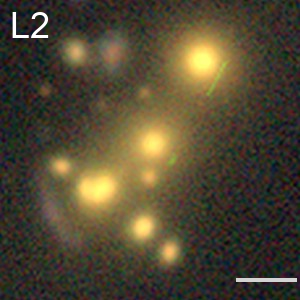}  
    \includegraphics[width=0.48\columnwidth]{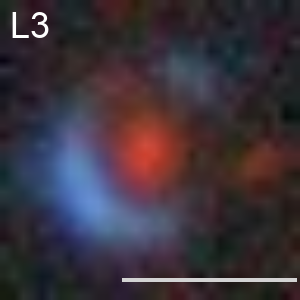}  
    \includegraphics[width=0.48\columnwidth]{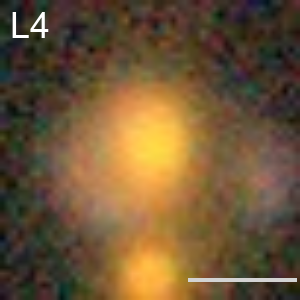}  
    \includegraphics[width=0.48\columnwidth]{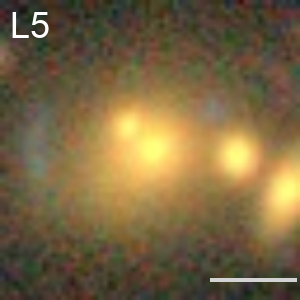} 
    \includegraphics[width=0.48\columnwidth]{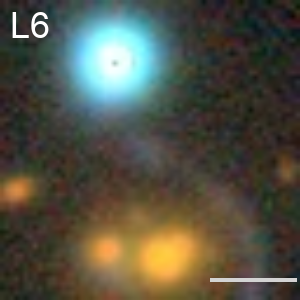}  
    \includegraphics[width=0.48\columnwidth]{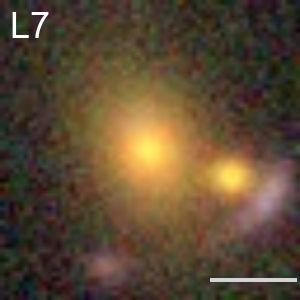}     
    \includegraphics[width=0.48\columnwidth]{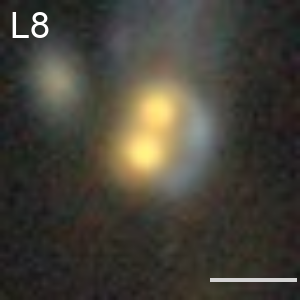}  
    \vspace{6pt}
    \caption{The strong gravitational lens candidates detected in the top 10 000 anomalies. More information about these sources can be found in Table \ref{tab: lens candidates}. The angular scale bar within each image represents 5 arcsec.}
    \label{fig: Anomalies lenses} 
\end{minipage}
\end{figure}

Fig.  \ref{fig: Anomalies lenses} and Table \ref{tab: lens candidates} show the strong gravitational lens candidates that have been identified in the top 10 000 anomalies. The source labelled L1 has been cross-matched, identified and confirmed to be a strong gravitational lens. The sources L2 through L5 have been cross-matched with other catalogues and identified as strong lens candidates due to the combined strong lens catalogued created in Grespan et al. (in preperation). The remaining sources are suspected strong lenses based on visual characteristics within the cut-outs, but have not been listed in any known catalogue. It should be noted that only fairly obvious lenses were labelled as interesting and there may be many more candidates in the list of anomalies that could potentially be identified by an expert in the field. 

Confirming the nature of these sources is challenging without significant additional analysis and spectroscopic follow-up. We found that the photometric redshift information available to us does not appear to be particularly reliable for these lensed systems. For instance, the DECaLS photometric redshifts for the sources in L1, which is a confirmed lens system, places the lens at a higher redshift than the background source, which is obviously incorrect. Of particular interest would be to confirm if the system U6 of Fig.  \ref{fig: Anomalies unusual} is actually a lensed system and if the two sources in the foreground are coincidental or in fact part of the lens. Similarly, the pair of sources in L8 would need spectroscopic information to determine if they are both involved in lensing the background source or not.

\begin{table}
\begin{centering}
    \caption{The gravitational lens candidates that have been identified in the top 10 000 anomalies. The second and third columns show the right ascension and declination in degrees, respectively. The last column indicates whether the candidates have been confirmed to be a lens, a lens candidate or a candidate that has not been matched to any other catalogue yet.}
    \label{tab: lens candidates}
    \begin{tabular*}{\columnwidth}{@{}l@{\hspace*{12pt}}c@{\hspace*{12pt}}c@{\hspace*{12pt}}c@{}}
        \hline
        \Tstrut Entry & RA & \textrm{Declination} & Information \\
            & [deg] & \textrm{[deg]} & \\
        [3pt] 
        \hline
        \Tstrut
        L1 &  35.2352 &  $-7.7199$ & Confirmed lens -- [\citet{2012ApJ...749...38M}] \\
        L2 &  27.9503 & $-32.6199$ & Candidate -- [\citet{2019ApJS..243...17J}] \\
        L3 &  61.6016 & $-26.7733$ & Candidate -- [\citet{2019ApJS..243...17J}] \\
        L4 & 128.1546 &  $13.5797$ & Candidate -- [\citet{2017ApJ...851...48S}] \\
        L5 & 340.2492 & $-52.7542$ & Candidate -- [\citet{2017ApJS..232...15D}] \\
        L6 &  78.3564 & $-30.8416$ & Candidate -- Previously undetected \\
        L7 &   9.7307 &   $7.3230$ & Candidate -- Previously undetected \\
        L8 &  60.1041 & $-16.3973$ & Candidate -- Previously undetected \\
        [3pt]
        \hline
    \end{tabular*}
    \end{centering}
\end{table}

\subsubsection{Galaxy merger candidates}
\label{Results: Galaxy Merger Candidates}

\begin{table}
    \caption{Coordinates of the galaxy merger candidates shown in Fig.  \ref{fig: Anomalies mergers}. The \textit{Entry} columns indicate the corresponding image as shown in Fig.  \ref{fig: Anomalies mergers} and the following columns show the right ascension and declination in degrees, respectively.}
    \label{tab: merger candidates}
        \begin{tabular*}{\columnwidth}{@{}l@{\hspace*{14pt}}c@{\hspace*{14pt}}c@{\hspace*{14pt}}c@{\hspace*{14pt}}c@{\hspace*{14pt}}c}
            \hline
            \Tstrut
            Entry & RA & \textrm{Declination} & Entry & RA & \textrm{Declination}\\
                & [deg] & \textrm{[deg]} & & [deg] & \textrm{[deg]} \\
            [3pt]
            \hline 
            \Tstrut
            M1  &  11.4865 &  $32.2940$ &     M25 & 333.0850 &   $0.5605$ \\
            M2  &  12.0642 & $-25.6886$ &     M26 & 333.2532 &  $21.9719$ \\
            M3  &  12.4649 &  $17.7756$ &     M27 & 335.2230 &  $13.4425$ \\
            M4  & 123.6086 &  $37.2613$ &     M28 &  36.9431 &  $26.5896$ \\
            M5  & 127.6318 &  $18.2050$ &     M29 &  40.4523 & $-49.0052$ \\
            M6  & 134.2013 &  $30.8611$ &     M30 &  42.9094 & $-16.6571$ \\
            M7  & 138.9657 &  $13.5864$ &     M31 &  43.6074 & $-64.1671$ \\
            M8  & 147.5532 &  $-5.6929$ &     M32 &  44.8704 & $-14.2910$ \\
            M9  & 168.9463 &  $15.8239$ &     M33 &  46.3564 & $-19.4730$ \\
            M10 & 192.1130 &  $15.5824$ &     M34 &  49.0504 & $-12.1633$ \\
            M11 & 205.4201 &  $13.5041$ &     M35 &  50.3167 & $-28.3164$ \\
            M12 & 213.6448 &  $24.4104$ &     M36 &  60.4281 & $-23.5613$ \\
            M13 &  22.8853 & $-22.3706$ &     M37 &  63.3138 & $-14.6525$ \\
            M14 & 222.3588 &  $23.3494$ &     M38 &  64.5453 & $-31.2488$ \\
            M15 & 229.9712 &   $2.6200$ &     M39 &  68.7329 & $-40.0342$ \\
            M16 & 239.7994 &  $20.7477$ &     M40 &  70.2079 & $-44.9443$ \\
            M17 &  24.0000 & $-11.7047$ &     M41 &  72.2531 & $-33.3201$ \\
            M18 &  25.4929 & $-13.8470$ &     M42 &  76.0434 & $-57.2689$ \\
            M19 & 251.9828 &  $10.5277$ &     M43 &  79.3557 & $-48.3955$ \\
            M20 & 254.6314 &  $58.9370$ &     M44 &  80.1525 & $-38.6625$ \\
            M21 &  28.6783 &  $27.3285$ &     M45 &  81.0061 & $-32.2876$ \\
            M22 &  30.6765 &  $-2.5885$ &     M46 &  81.8048 & $-31.4080$ \\
            M23 & 318.6995 &  $-2.1876$ &     M47 &  83.3110 & $-39.4453$ \\
            M24 & 330.9685 &  $-1.6943$ &     M48 &  85.2283 & $-20.3304$ \\
            [2pt]
            \hline
        \end{tabular*}
\end{table}

\begin{figure*}
    \includegraphics[width=0.328\columnwidth]{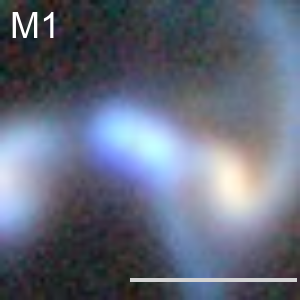}\hfill
    \includegraphics[width=0.328\columnwidth]{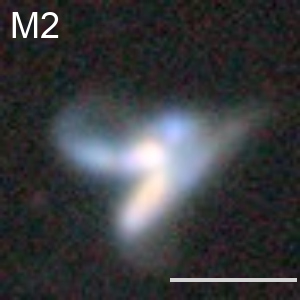}\hfill
    \includegraphics[width=0.328\columnwidth]{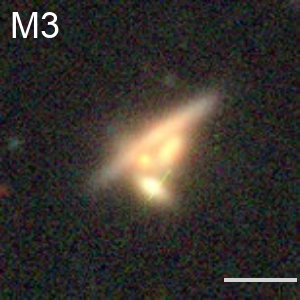}\hfill
    \includegraphics[width=0.328\columnwidth]{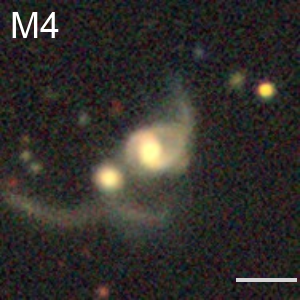}\hfill
    \includegraphics[width=0.328\columnwidth]{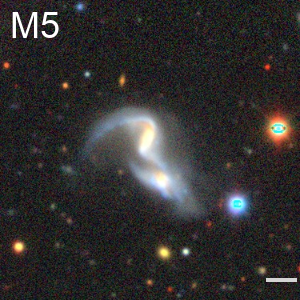}\hfill
    \includegraphics[width=0.328\columnwidth]{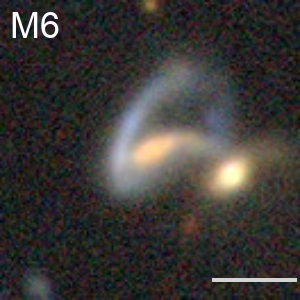}\hfill
    \includegraphics[width=0.328\columnwidth]{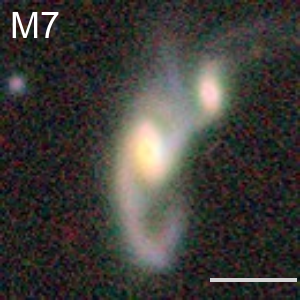}\hfill
    \includegraphics[width=0.328\columnwidth]{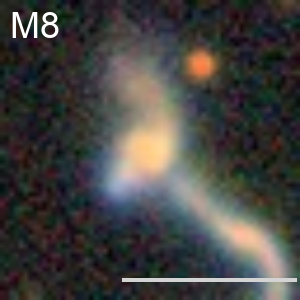}\hfill
    \includegraphics[width=0.328\columnwidth]{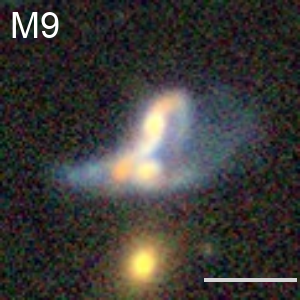}\hfill
    \includegraphics[width=0.328\columnwidth]{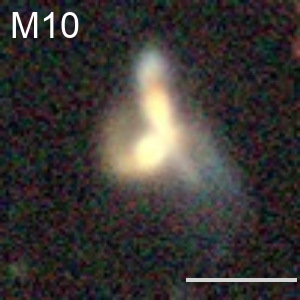}\hfill
    \includegraphics[width=0.328\columnwidth]{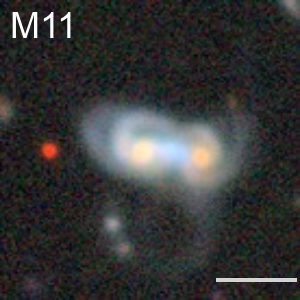}\hfill
    \includegraphics[width=0.328\columnwidth]{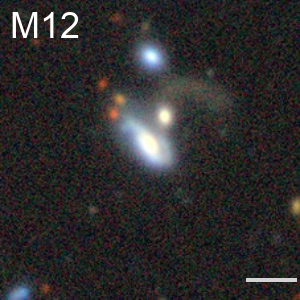}\hfill
    \includegraphics[width=0.328\columnwidth]{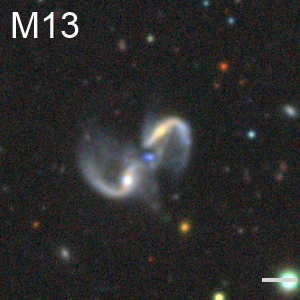}\hfill
    \includegraphics[width=0.328\columnwidth]{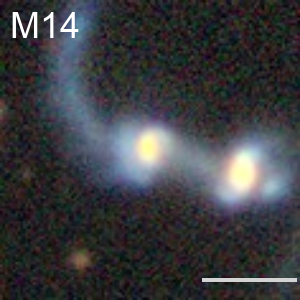}\hfill
    \includegraphics[width=0.328\columnwidth]{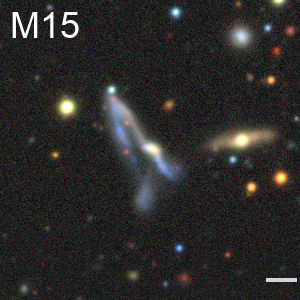}\hfill
    \includegraphics[width=0.328\columnwidth]{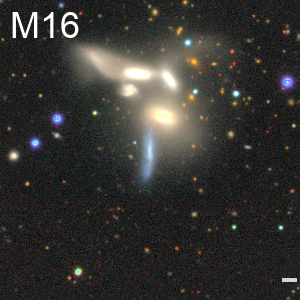}\hfill
    \includegraphics[width=0.328\columnwidth]{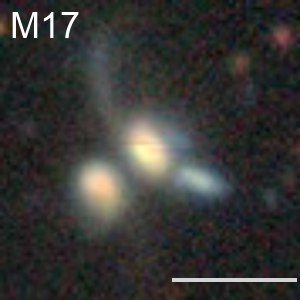}\hfill
    \includegraphics[width=0.328\columnwidth]{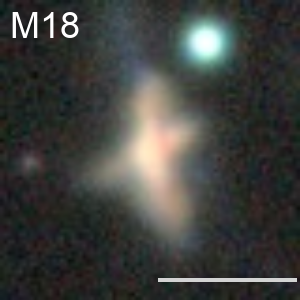}\hfill
    \includegraphics[width=0.328\columnwidth]{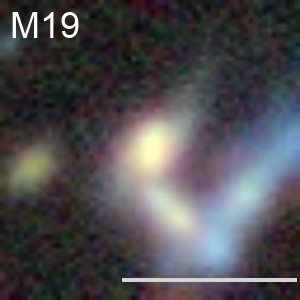}\hfill
    \includegraphics[width=0.328\columnwidth]{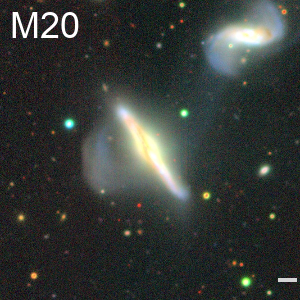}\hfill
    \includegraphics[width=0.328\columnwidth]{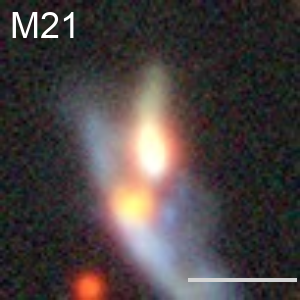}\hfill
    \includegraphics[width=0.328\columnwidth]{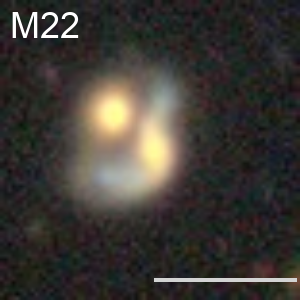}\hfill 
    \includegraphics[width=0.328\columnwidth]{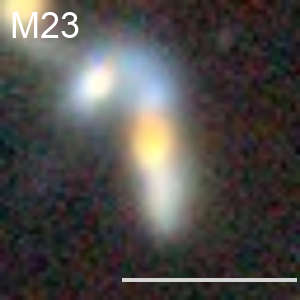}\hfill
    \includegraphics[width=0.328\columnwidth]{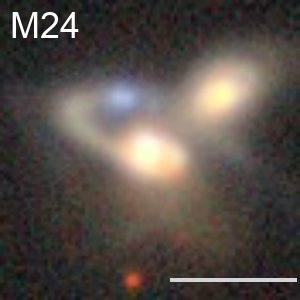}\hfill
    \includegraphics[width=0.328\columnwidth]{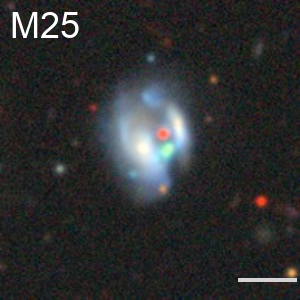}\hfill
    \includegraphics[width=0.328\columnwidth]{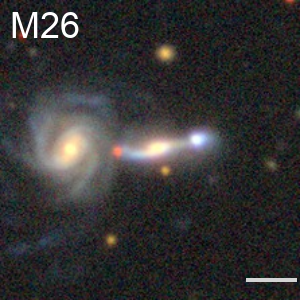}\hfill
    \includegraphics[width=0.328\columnwidth]{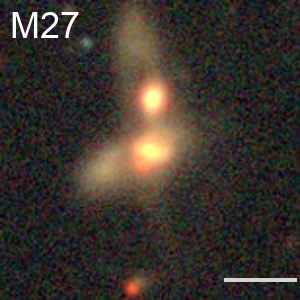}\hfill
    \includegraphics[width=0.328\columnwidth]{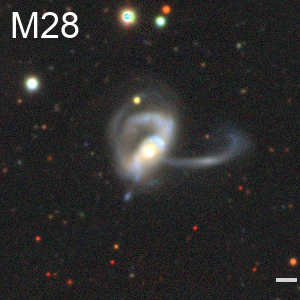}\hfill
    \includegraphics[width=0.328\columnwidth]{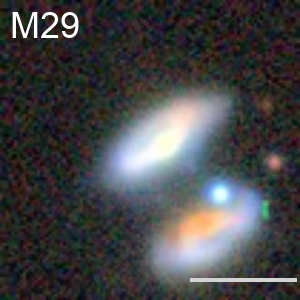}\hfill
    \includegraphics[width=0.328\columnwidth]{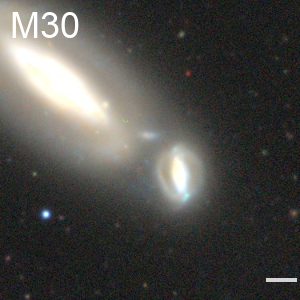}\hfill
    \includegraphics[width=0.328\columnwidth]{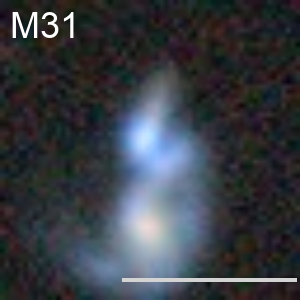}\hfill
    \includegraphics[width=0.328\columnwidth]{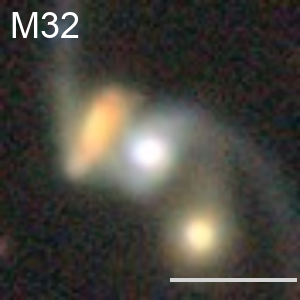}\hfill
    \includegraphics[width=0.328\columnwidth]{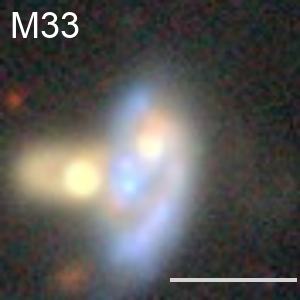}\hfill
    \includegraphics[width=0.328\columnwidth]{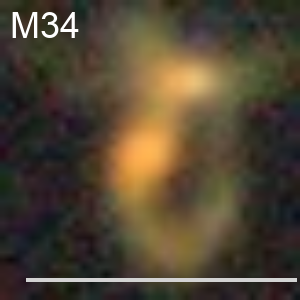}\hfill
    \includegraphics[width=0.328\columnwidth]{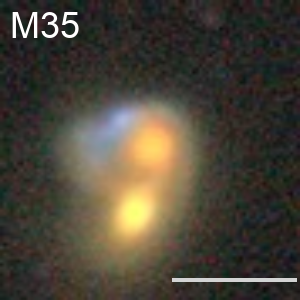}\hfill
    \includegraphics[width=0.328\columnwidth]{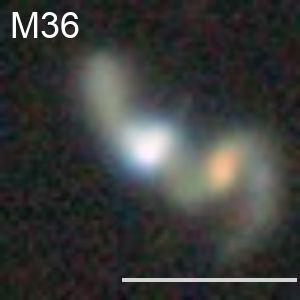}\hfill
    \includegraphics[width=0.328\columnwidth]{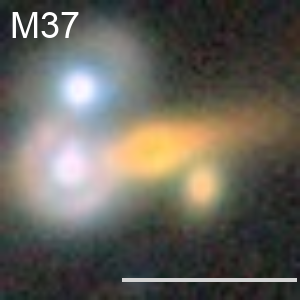}\hfill
    \includegraphics[width=0.328\columnwidth]{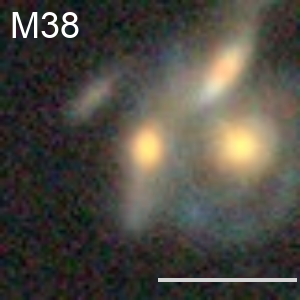}\hfill
    \includegraphics[width=0.328\columnwidth]{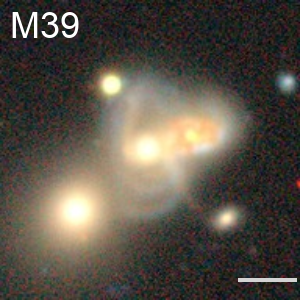}\hfill
    \includegraphics[width=0.328\columnwidth]{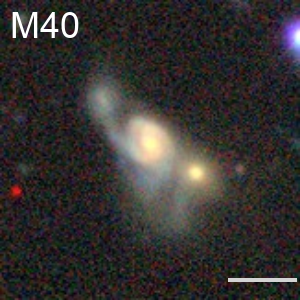}\hfill
    \includegraphics[width=0.328\columnwidth]{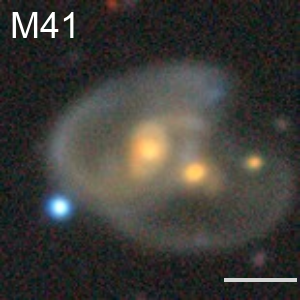}\hfill
    \includegraphics[width=0.328\columnwidth]{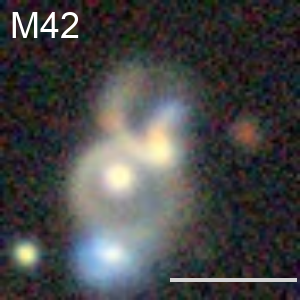}\hfill
    \includegraphics[width=0.328\columnwidth]{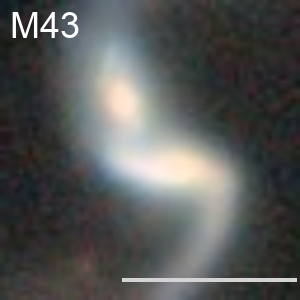}\hfill
    \includegraphics[width=0.328\columnwidth]{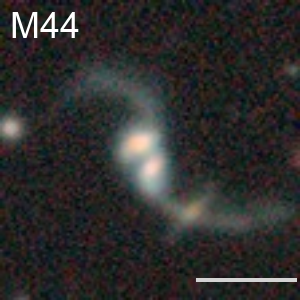}\hfill
    \includegraphics[width=0.328\columnwidth]{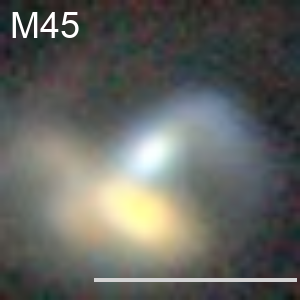}\hfill
    \includegraphics[width=0.328\columnwidth]{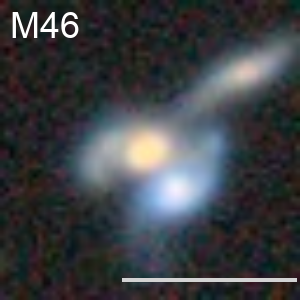}\hfill
    \includegraphics[width=0.328\columnwidth]{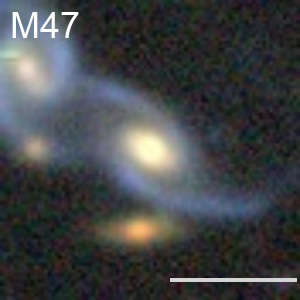}\hfill
    \includegraphics[width=0.328\columnwidth]{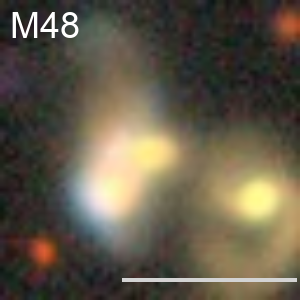}\hfill \\
    \caption{Some of the galaxy merger candidates that were found in the top 10 000 of the main set. The images displayed are those that were visually the most interesting out of the 1609 merger candidates. The angular scale bar within each image represents 10 arcsec. More information can be found in Table \ref{tab: merger candidates}.}
    \label{fig: Anomalies mergers}
\end{figure*}

Fig.  \ref{fig: Anomalies mergers} shows some of the more striking examples of mergers detected with \astronomaly{}. The 1609 galaxy merger candidates were compared with the Catalog of Morphologically Identified Merging Galaxies (Hwang \& Chang \citeyear{Hwang2009}), but the sky coverage of this catalogue, 422 deg$^2$, is significantly smaller than that of DECaLS, 14 000 deg$^2$, so not many matches were expected. In addition, the different data cuts applied make a direct comparison challenging. However, six matches were found at a cross-matching distance of 30 arcsec. These six sources were visually confirmed to match the sources in the catalogue. While spectroscopic follow-up would be needed to further investigate the rest of the sources shown in Fig.  \ref{fig: Anomalies mergers}, the presence of significant tidal streams and interaction suggests the majority of these are high-probability merger candidates. 

These results show that \astronomaly{} could be used to build a large, albeit incomplete, merger catalogue. To gauge the theoretical completeness of such a catalogue, we compared the precision values for the identified merger candidates from our evaluation subset, first described in Section \ref{Data: Finding An Evaluation Set}, with a supervised method from O’Ryan et al. \citeyearpar{O_Ryan_2023}, which used the same CNN to classify mergers. It is important to note that O'Ryan et al. used data from the European Space Agency's Hubble Space Telescope Science Archive\footnote{See \url{http://hst.esac.esa.int/ehst/}}, which has significantly higher data quality compared to DECaLS. Unfortunately, as no similar work has been published on DECaLS data, this was the best comparison that could be made at this time. Deriving conclusions from comparisons in performance metrics should therefore be done cautiously because of the discrepancy in data quality. 

The entire main subset was initially ordered by the `predicted\_user\_score', discussed in Section \ref{Methodology: Anomaly Detection}, which we found best identified merger candidates for manual labelling. For more details on the scores implemented in \astronomaly{}, refer to Lochner \& Bassett \citeyearpar{Lochner_2021}. This ordering can be considered the equivalent of a classifier probability, allowing the application of different thresholds to `classify' a source as a merger. We then extracted the index (position in the ordered list) for the evaluation subset, for which we have labels. We considered a merger candidate to be the positive class, and all others (including other genuine anomalies) as the negative class. By applying different thresholds in the index, we can obtain candidate merger samples of differing purity. To compare with O’Ryan et al. \citeyearpar{O_Ryan_2023}, we selected thresholds resulting in specified recall values and report the corresponding precision in Table \ref{tab: precision comparison}.

Our unsupervised method excelled with high precision up to a 50 per cent recall threshold, outperforming the supervised method. However, the supervised method demonstrated superior precision at higher recall thresholds, emphasizing its strength in completeness. Our method's higher initial precision makes it well suited for efficiently obtaining a sample of desired sources. This subset can then serve as valuable training data for a subsequent supervised method.

In addition, as mergers make up the majority of anomalous sources, one could explore the application of either a second round of AL or a trained classifier to actively remove them, which would help highlight more unusual sources.

\begin{table}
\begin{centering}
    \caption{Comparison of precision values based on predicted user score with a supervised method implemented by O’Ryan et al. \citeyearpar{O_Ryan_2023}. Precision values are calculated at the various specified recall thresholds.}
    \label{tab: precision comparison}
        \begin{tabular*}{\columnwidth}{@{}l@{\hspace*{8pt}}c@{\hspace*{8pt}}c@{\hspace*{8pt}}c@{\hspace*{8pt}}c@{}}
        \hline
        \Tstrut Recall thresholds & 25 per cent & 50 per cent & 75 per cent & 95 per cent \\
        [3pt] 
        \hline
        \Tstrut Supervised Method & 0.939 & 0.877 & 0.781 & 0.538 \\
        Predicted User Score & 1.0 & 0.982 & 0.584 & 0.353 \\
        [2pt] 
        \hline
        \end{tabular*}
    \end{centering}
\end{table}

\vspace{-10pt}

\section{Conclusions}
\label{Conclusions}
Anomaly detection on a large scale is critical for scientific discovery in current and future astronomical data sets. Computational challenges are abundant and more often than not, supervised methods are used to detect ``new'' sources that are of the same type as ones already identified. The search for novel anomalies requires unsupervised methods and has previously been applied on relatively modest scales using frameworks like \astronomaly{}. In this work, we have applied \astronomaly{} to a much larger scale: 3 884 404 galaxies obtained from the DECaLS DR8. This was a test to determine if these methods could be applied on significantly larger scales and if they could find interesting sources in the relatively unexplored DECaLS data set.

We looked at various options for selecting a DECaLS subset to study and discovered that the choice of data selection criteria had a significant impact on the balance between scalability and discovery, which is important for detecting anomalies. If the selection criteria incorporated are too strict, anomalies might be overlooked. The selection criteria we used impacted the number of gravitational lens candidates identified within our evaluation set and would consequently constrain the number of such candidates within the larger data set used. We were unable to identify a set of cuts that could reduce the data set size while retaining all the lens candidates.  Exploring criteria that have less impact on the count of lens candidates could be a worthwhile avenue for future research. To ensure a thorough investigation, data should be as wide and free as possible, but should still remain manageable enough in size to analyse. 

As the size of the data set increases, the challenges of large data volumes, storage, and computational complexity become more important. One of the main challenges that we experienced was the transfer of data from the host server to a local computer, which took several weeks as the data had to be validated too. As the size and complexity of data sets increase, such as in the case of the LSST and the SKA, the traditional approach of moving the data to the compute nodes becomes impractical and costly. Bringing the compute to the data is therefore essential for anomaly detection in big data sets. 

Pre-processing can also have a significant impact on the anomalies detected. Following Walmsley et al. \citeyearpar{Walmsley_2021}, we greyscaled the images before feature extraction through a straightforward averaging of the three optical channels. However, it is worth noting that using established band weightings, such as those available in OpenCV\footnote{\url{https://docs.opencv.org/4.x/de/d25/imgproc_color_conversions.html}}, could potentially offer a more suitable approach for creating the greyscale images (Etsebeth \citeyear{Etsebeth}). Furthermore, our use of the basic sigma clipping procedure, as described in Walmsley et al. \citeyearpar{Walmsley_2021}, proved effective in reducing noise but at times could not accommodate nearby bright sources within the cut-outs. Alternative techniques, like the sigma clipping and masking procedure used in Lochner \& Bassett \citeyearpar{Lochner_2021}, may provide improved results by ensuring the removal of all sources that are not part of the target object.

For feature extraction, we showed that using the pre-trained CNN from Walmsley et al. \citeyearpar{Walmsley_2021} to obtain representations of images works extremely well for DECaLS data, further demonstrating the value of CNNs as general-purpose feature extractors. This adaptability reduces the need for expensive labelling and allows the reuse of networks for other, unsupervised tasks. It also highlights the value of training networks to solve complex classification tasks on large data sets and publicly releasing the trained network weights and code for use by the community on other tasks and data sets.


We applied \astronomaly{} using the anomaly detection algorithm, iForest, to the features extracted using the CNN. iForest scaled well to the considerable amount of data we had and was able to effectively identify artefacts, which are anomalies but are not scientifically interesting. The fact that a large number of artefacts were present in the list of anomalies, despite our selection cuts, highlights the difficulty of detecting artefacts with automated flagging techniques but also the value of unsupervised methods in detecting artefacts that were missed. However, iForest alone could not differentiate between interesting and less interesting anomalies. We found that AL, using a relatively small amount of human labelling, is critical to success in anomaly detection for large, uncurated astronomical data sets. Using \astronomaly{}'s interface, it only took one person a few hours to label enough data to enable significant discoveries.

Fig.  \ref{fig: Anomalies in top 2000: labelled} shows that AL algorithms can be used to enhance the performance of anomaly detection algorithms by prioritizing the most relevant anomalies and filtering out less interesting ones. We compared two different AL approaches and found similar performance (Fig.  \ref{fig: AL vs GP for all anomalies}). We chose the NS due to computational challenges with the direct regression method at a larger scale. While more computationally efficient GP methods are available, the tests on the evaluation set suggested that the extra implementation effort might not justify potential performance gains. This may not be the case for other data sets (e.g. Walmsley et al. \citeyear{Walmsley_2021}).

The application of anomaly detection, in conjunction with the NS active learning method using a total of 10 000 labels, on the data set of 3 884 404 sources, identified 1635 interesting anomalies in the top 2000 sources. Of these, eight gravitational lens candidates were identified, five of which are listed as candidates in other catalogues. In addition, 1609 sources were identified that contain galaxies exhibiting some signs of a gravitational merger event. We compared \astronomaly{}'s ability to detect merger candidates with that of a supervised method. Unsurprisingly, the supervised approach outperformed the unsupervised method in completeness but we did find our method was able to more rapidly recover a pure sample of merger candidates, which could then be used as a training set for subsequent classifiers. Unfortunately the number of lenses detected was too few to perform a similar analysis but it is not unreasonable to assume that, with enough examples, \astronomaly{} could also be effective at finding an initial sample of lenses or other sources with unusual morphology.

Finally, 18 sources, shown in Fig.  \ref{fig: Anomalies unusual}, were found that were unstudied to the best of our current knowledge. These unusual sources vary in morphology and require additional investigation in order to identify their nature. They include ring galaxies exhibiting strange colours and morphology, a source that is half red and half blue, a potentially strong lensed system with a pair of sources acting as the lens, several known interacting groups and some sources that are either interacting or coincidental alignments. Moreover, it is important to note, that these sources were all contained within only the top 2000 most anomalous sources after applying 10 000 labels. This opens up the potential for significantly more sources that are also interesting to be identified. 

Our results show that the modern anomaly detection techniques included in \astronomaly{} scale well to large data sets and are capable of rapidly detecting scientifically interesting anomalies. As the number and quality of anomalies detected can be affected by selection cuts, these should be avoided as far as possible by leveraging computationally intensive unsupervised frameworks running on remote data centres. This work paves the way for scientific discovery with anomaly detection in large data sets, such as those expected from the Vera C. Rubin Observatory, Euclid and the Square Kilometre Array.

\section*{Acknowledgements}
\label{Acknowledgements}
The authors would like to personally thank Aritra Ghosh for the input into identifying the unusual anomalies detected in this work. We would also like to thank the anonymous referee for their helpful comments which improved the draft.

VE and ML acknowledge support from the South African Radio Astronomy Observatory and the National Research Foundation (NRF) towards this research. Opinions expressed and conclusions arrived at are those of the authors and are not necessarily to be attributed to the NRF.

MW is a Dunlap Fellow and acknowledges funding from the Science and Technology Facilities Council (STFC) Grant Code ST/R505006/1.

This paper includes data that have been provided by AAO Data Central (datacentral.org.au).

The Legacy Surveys consist of three individual and complementary projects: the Dark Energy Camera Legacy Survey (DECaLS; Proposal ID \#2014B-0404; PIs: David Schlegel and Arjun Dey), the Beijing--Arizona Sky Survey (BASS; NOAO Prop. ID \#2015A-0801; PIs: Zhou Xu and Xiaohui Fan), and the Mayall $z$-band Legacy Survey (MzLS; Prop. ID \#2016A-0453; PI: Arjun Dey). DECaLS, BASS, and MzLS together include data obtained, respectively, at the Blanco telescope, Cerro Tololo Inter-American Observatory, NSF’s NOIRLab; the Bok telescope, Steward Observatory, University of Arizona; and the Mayall telescope, Kitt Peak National Observatory, NOIRLab. Pipeline processing and analyses of the data were supported by NOIRLab and the Lawrence Berkeley National Laboratory (LBNL). The Legacy Surveys project is honoured to be permitted to conduct astronomical research on Iolkam Du’ag (Kitt Peak), a mountain with particular significance to the Tohono O’odham Nation.

NOIRLab is operated by the Association of Universities for Research in Astronomy (AURA) under a cooperative agreement with the National Science Foundation. LBNL is managed by the Regents of the University of California under contract to the U.S. Department of Energy.

This project used data obtained with the Dark Energy Camera (DECam), which was constructed by the Dark Energy Survey (DES) collaboration. Funding for the DES Projects has been provided by the U.S. Department of Energy, the U.S. National Science Foundation, the Ministry of Science and Education of Spain, the Science and Technology Facilities Council of the United Kingdom, the Higher Education Funding Council for England, the National Center for Supercomputing Applications at the University of Illinois at Urbana-Champaign, the Kavli Institute of Cosmological Physics at the University of Chicago, Center for Cosmology and Astro-Particle Physics at the Ohio State University, the Mitchell Institute for Fundamental Physics and Astronomy at Texas A\&M University, Financiadora de Estudos e Projetos, Fundacao Carlos Chagas Filho de Amparo, Financiadora de Estudos e Projetos, Fundacao Carlos Chagas Filho de Amparo a Pesquisa do Estado do Rio de Janeiro, Conselho Nacional de Desenvolvimento Cientifico e Tecnologico and the Ministerio da Ciencia, Tecnologia e Inovacao, the Deutsche Forschungsgemeinschaft and the Collaborating Institutions in the Dark Energy Survey. The Collaborating Institutions are Argonne National Laboratory, the University of California at Santa Cruz, the University of Cambridge, Centro de Investigaciones Energeticas, Medioambientales y Tecnologicas-Madrid, the University of Chicago, University College London, the DES-Brazil Consortium, the University of Edinburgh, the Eidgenossische Technische Hochschule (ETH) Zurich, Fermi National Accelerator Laboratory, the University of Illinois at Urbana-Champaign, the Institut de Ciencies de l’Espai (IEEC/CSIC), the Institut de Fisica d’Altes Energies, Lawrence Berkeley National Laboratory, the Ludwig Maximilians Universitat Munchen and the associated Excellence Cluster Universe, the University of Michigan, NSF’s NOIRLab, the University of Nottingham, the Ohio State University, the University of Pennsylvania, the University of Portsmouth, SLAC National Accelerator Laboratory, Stanford University, the University of Sussex, and Texas A\&M University.

BASS is a key project of the Telescope Access Program (TAP), which has been funded by the National Astronomical Observatories of China, the Chinese Academy of Sciences (the Strategic Priority Research Program “The Emergence of Cosmological Structures” Grant \# XDB09000000), and the Special Fund for Astronomy from the Ministry of Finance. The BASS is also supported by the External Cooperation Program of Chinese Academy of Sciences (Grant \# 114A11KYSB20160057), and Chinese National Natural Science Foundation (Grant \# 12120101003, \# 11433005).

The Legacy Survey team uses data products from the Near-Earth Object Wide-field Infrared Survey Explorer (NEOWISE), which is a project of the Jet Propulsion Laboratory/California Institute of Technology. NEOWISE is funded by the National Aeronautics and Space Administration.

The Legacy Surveys imaging of the DESI footprint is supported by the Director, Office of Science, Office of High Energy Physics of the U.S. Department of Energy under Contract No. DE-AC02-05CH1123, by the National Energy Research Scientific Computing Center, a DOE Office of Science User Facility under the same contract; and by the U.S. National Science Foundation, Division of Astronomical Sciences under Contract No. AST-0950945 to NOAO.

We acknowledge the use of the ilifu cloud computing facility -– www.ilifu.ac.za, a partnership between the University of Cape Town, the University of the Western Cape, the University of Stellenbosch, Sol Plaatje University and the Cape Peninsula University of Technology. The Ilifu facility is supported by contributions from the Inter-University Institute for Data Intensive Astronomy (IDIA) -– a partnership between the University of Cape Town, the University of Pretoria, the University of the Western Cape, the Computational Biology division at UCT and the Data Intensive Research Initiative of South Africa (DIRISA).

\section*{Data Availability}
\label{Data Availability}
The data used in this paper from the DECaLS survey are publicly available. Catalogues of potential merger candidates and other types of sources can be made available on request to the authors.

\bibliographystyle{mnras}
\bibliography{bibliography} 

\begin{thebibliography}{}
\makeatletter
\relax
\def\mn@urlcharsother{\let\do\@makeother \do\$\do\&\do\#\do\^\do\_\do\%\do\~}
\def\mn@doi{\begingroup\mn@urlcharsother \@ifnextchar [ {\mn@doi@} {\mn@doi@[]}}
\def\mn@doi@[#1]#2{\def\@tempa{#1}\ifx\@tempa\@empty \href {http://dx.doi.org/#2} {doi:#2}\else \href {http://dx.doi.org/#2} {#1}\fi \endgroup}
\def\mn@eprint#1#2{\mn@eprint@#1:#2::\@nil}
\def\mn@eprint@arXiv#1{\href {http://arxiv.org/abs/#1} {{\tt arXiv:#1}}}
\def\mn@eprint@dblp#1{\href {http://dblp.uni-trier.de/rec/bibtex/#1.xml} {dblp:#1}}
\def\mn@eprint@#1:#2:#3:#4\@nil{\def\@tempa {#1}\def\@tempb {#2}\def\@tempc {#3}\ifx \@tempc \@empty \let \@tempc \@tempb \let \@tempb \@tempa \fi \ifx \@tempb \@empty \def\@tempb {arXiv}\fi \@ifundefined {mn@eprint@\@tempb}{\@tempb:\@tempc}{\expandafter \expandafter \csname mn@eprint@\@tempb\endcsname \expandafter{\@tempc}}}

\bibitem[\protect\citeauthoryear{{Ahumada} et~al.,}{{Ahumada} et~al.}{2020}]{SDSS2020}
{Ahumada} R.,  et~al., 2020, \mn@doi [\apjs] {10.3847/1538-4365/ab929e}, \href {https://ui.adsabs.harvard.edu/abs/2020ApJS..249....3A} {249, 3}

\bibitem[\protect\citeauthoryear{{Almeida} et~al.,}{{Almeida} et~al.}{2023}]{almeida2023eighteenth}
{Almeida} A.,  et~al., 2023, \mn@doi [\apjs] {10.3847/1538-4365/acda98}, \href {https://ui.adsabs.harvard.edu/abs/2023ApJS..267...44A} {267, 44}

\bibitem[\protect\citeauthoryear{{Baron} \& {Poznanski}}{{Baron} \& {Poznanski}}{2017}]{2017MNRAS.465.4530B}
{Baron} D.,  {Poznanski} D.,  2017, \mn@doi [\mnras] {10.1093/mnras/stw3021}, \href {https://ui.adsabs.harvard.edu/abs/2017MNRAS.465.4530B} {465, 4530}

\bibitem[\protect\citeauthoryear{Breiman}{Breiman}{2001}]{Breiman2001}
Breiman L.,  2001, \mn@doi [Machine Learn.] {10.1023/a:1010933404324}, 45, 5

\bibitem[\protect\citeauthoryear{Breunig, Kriegel, Ng  \& Sander}{Breunig et~al.}{2000}]{10.1145/335191.335388}
Breunig M.~M.,  Kriegel H.-P.,  Ng R.~T.,   Sander J.,  2000, \mn@doi [ACM SIGMOD Rec.] {10.1145/335191.335388}, 29, 93–104

\bibitem[\protect\citeauthoryear{{Chambers} et~al.,}{{Chambers} et~al.}{2016}]{Chambers2016}
{Chambers} K.~C.,  et~al., 2016, \mn@doi [\pasp] {10.1088/1538-3873/128/968/104502}, \href {https://ui.adsabs.harvard.edu/abs/2016PASP..128j4502C} {128, 104502}

\bibitem[\protect\citeauthoryear{{Condon}, {Cotton}, {Greisen}, {Yin}, {Perley}, {Taylor}  \& {Broderick}}{{Condon} et~al.}{1998}]{1998AJ....115.1693C}
{Condon} J.~J.,  {Cotton} W.~D.,  {Greisen} E.~W.,  {Yin} Q.~F.,  {Perley} R.~A.,  {Taylor} G.~B.,   {Broderick} J.~J.,  1998, \mn@doi [\aj] {10.1086/300337}, \href {https://ui.adsabs.harvard.edu/abs/1998AJ....115.1693C} {115, 1693}

\bibitem[\protect\citeauthoryear{{D{\'a}lya} et~al.,}{{D{\'a}lya} et~al.}{2018}]{GLADE2018}
{D{\'a}lya} G.,  et~al., 2018, \mn@doi [\mnras] {10.1093/mnras/sty1703}, \href {https://ui.adsabs.harvard.edu/abs/2018MNRAS.479.2374D} {479, 2374}

\bibitem[\protect\citeauthoryear{Debosscher, Sarro, Aerts, Cuypers, Vandenbussche, Garrido  \& Solano}{Debosscher et~al.}{2007}]{Debosscher_2007}
Debosscher J.,  Sarro L.~M.,  Aerts C.,  Cuypers J.,  Vandenbussche B.,  Garrido R.,   Solano E.,  2007, \mn@doi [A \& A] {10.1051/0004-6361:20077638}, 475, 1159

\bibitem[\protect\citeauthoryear{Dey et~al.,}{Dey et~al.}{2019}]{Dey_2019}
Dey A.,  et~al., 2019, \mn@doi [AJ] {10.3847/1538-3881/ab089d}, 157, 168

\bibitem[\protect\citeauthoryear{{Diehl} et~al.,}{{Diehl} et~al.}{2017}]{2017ApJS..232...15D}
{Diehl} H.~T.,  et~al., 2017, \mn@doi [\apjs] {10.3847/1538-4365/aa8667}, \href {https://ui.adsabs.harvard.edu/abs/2017ApJS..232...15D} {232, 15}

\bibitem[\protect\citeauthoryear{{Eke} et~al.,}{{Eke} et~al.}{2004}]{Eke2004}
{Eke} V.~R.,  et~al., 2004, \mn@doi [\mnras] {10.1111/j.1365-2966.2004.07408.x}, \href {https://ui.adsabs.harvard.edu/abs/2004MNRAS.348..866E} {348, 866}

\bibitem[\protect\citeauthoryear{Etsebeth}{Etsebeth}{2020}]{Etsebeth}
Etsebeth V.,  2020, Master's thesis, Univ. Western Cape, Cape Town, South Africa, \url {http://hdl.handle.net/11394/9028}

\bibitem[\protect\citeauthoryear{{Flesch}}{{Flesch}}{2023}]{Flesch2023}
{Flesch} E.~W.,  2023, \mn@doi [The Open Journal of Astrophys.] {10.21105/astro.2308.01505}, \href {https://ui.adsabs.harvard.edu/abs/2023OJAp....6E..49F} {6, 49}

\bibitem[\protect\citeauthoryear{{Giles} \& {Walkowicz}}{{Giles} \& {Walkowicz}}{2019}]{2019MNRAS.484..834G}
{Giles} D.,  {Walkowicz} L.,  2019, \mn@doi [\mnras] {10.1093/mnras/sty3461}, \href {https://ui.adsabs.harvard.edu/abs/2019MNRAS.484..834G} {484, 834}

\bibitem[\protect\citeauthoryear{Huang et~al.,}{Huang et~al.}{2020}]{Huang_2020}
Huang X.,  et~al., 2020, \mn@doi [ApJ] {10.3847/1538-4357/ab7ffb}, 894, 78

\bibitem[\protect\citeauthoryear{Huang, Storfer, Gu, Ravi, Pilon  \& et al}{Huang et~al.}{2021}]{Huang_2021}
Huang X.,  Storfer C.,  Gu A.,  Ravi V.,  Pilon A.,   et al 2021, \mn@doi [ApJ] {10.3847/1538-4357/abd62b}, 909, 27

\bibitem[\protect\citeauthoryear{Hwang \& Chang}{Hwang \& Chang}{2009}]{Hwang2009}
Hwang C.-Y.,  Chang M.-Y.,  2009, \mn@doi [ApJS] {10.1088/0067-0049/181/1/233}, 181, 233

\bibitem[\protect\citeauthoryear{{Ivezi{\'c}} et~al.,}{{Ivezi{\'c}} et~al.}{2019}]{2019ApJ...873..111I}
{Ivezi{\'c}} {\v{Z}}.,  et~al., 2019, \mn@doi [\apj] {10.3847/1538-4357/ab042c}, \href {https://ui.adsabs.harvard.edu/abs/2019ApJ...873..111I} {873, 111}

\bibitem[\protect\citeauthoryear{{Jacobs} et~al.,}{{Jacobs} et~al.}{2019}]{2019ApJS..243...17J}
{Jacobs} C.,  et~al., 2019, \mn@doi [\apjs] {10.3847/1538-4365/ab26b6}, \href {https://ui.adsabs.harvard.edu/abs/2019ApJS..243...17J} {243, 17}

\bibitem[\protect\citeauthoryear{{Jones} et~al.,}{{Jones} et~al.}{2009}]{6df2009}
{Jones} D.~H.,  et~al., 2009, \mn@doi [\mnras] {10.1111/j.1365-2966.2009.15338.x}, \href {https://ui.adsabs.harvard.edu/abs/2009MNRAS.399..683J} {399, 683}

\bibitem[\protect\citeauthoryear{Liaw \& Wiener}{Liaw \& Wiener}{2002}]{Randon_Forest}
Liaw A.,  Wiener M.,  2002, R News, 2, 18

\bibitem[\protect\citeauthoryear{{Lintott} et~al.,}{{Lintott} et~al.}{2008}]{lintott2008}
{Lintott} C.~J.,  et~al., 2008, \mn@doi [\mnras] {10.1111/j.1365-2966.2008.13689.x}, \href {https://ui.adsabs.harvard.edu/abs/2008MNRAS.389.1179L} {389, 1179}

\bibitem[\protect\citeauthoryear{{Lintott} et~al.,}{{Lintott} et~al.}{2011}]{lintott2011}
{Lintott} C.,  et~al., 2011, \mn@doi [\mnras] {10.1111/j.1365-2966.2010.17432.x}, \href {https://ui.adsabs.harvard.edu/abs/2011MNRAS.410..166L} {410, 166}

\bibitem[\protect\citeauthoryear{Liu, Ting  \& Zhou}{Liu et~al.}{2008}]{10.1109/ICDM.2008.17}
Liu F.~T.,  Ting K.~M.,   Zhou Z.-H.,  2008, in Bonchi F., ed, Proceedings of the 2008 Eighth IEEE International Conference on Data Mining. Icdm '08.
IEEE Computer Society, Usa, p. 413–422, \mn@doi{10.1109/icdm.2008.17}, \url {https://doi.org/10.1109/ICDM.2008.17}

\bibitem[\protect\citeauthoryear{Lochner \& Bassett}{Lochner \& Bassett}{2021}]{Lochner_2021}
Lochner M.,  Bassett B.,  2021, \mn@doi [Astron. Comput.] {10.1016/j.ascom.2021.100481}, 36, 100481

\bibitem[\protect\citeauthoryear{Mao, Geha, Wechsler, Weiner, Tollerud, Nadler  \& Kallivayalil}{Mao et~al.}{2021}]{Mao_2021}
Mao Y.-Y.,  Geha M.,  Wechsler R.~H.,  Weiner B.,  Tollerud E.~J.,  Nadler E.~O.,   Kallivayalil N.,  2021, \mn@doi [ApJ] {10.3847/1538-4357/abce58}, 907, 85

\bibitem[\protect\citeauthoryear{{Martin} et~al.,}{{Martin} et~al.}{2005}]{2005ApJ...619L...1M}
{Martin} D.~C.,  et~al., 2005, \mn@doi [\apjl] {10.1086/426387}, \href {https://ui.adsabs.harvard.edu/abs/2005ApJ...619L...1M} {619, L1}

\bibitem[\protect\citeauthoryear{{Martinazzo}, {Espadoto}  \& {Hirata}}{{Martinazzo} et~al.}{2020}]{cscv}
{Martinazzo} A.,  {Espadoto} M.,   {Hirata} N. S.~T.,  2020, in 25th International Conference on Pattern Recognition (ICPR). Italy, p.~4176, \url {https://ui.adsabs.harvard.edu/abs/2020arXiv200411336M}

\bibitem[\protect\citeauthoryear{Massey, Neugent  \& Levesque}{Massey et~al.}{2019}]{Massey_2019}
Massey P.,  Neugent K.~F.,   Levesque E.~M.,  2019, \mn@doi [AJ] {10.3847/1538-3881/ab1aa1}, 157, 227

\bibitem[\protect\citeauthoryear{{McInnes}, {Healy}  \& {Melville}}{{McInnes} et~al.}{2018}]{mcinnes2020umap}
{McInnes} L.,  {Healy} J.,   {Melville} J.,  2018, \mn@doi [arXiv e-prints] {10.48550/arXiv.1802.03426}, \href {https://ui.adsabs.harvard.edu/abs/2018arXiv180203426M} {p. arXiv:1802.03426}

\bibitem[\protect\citeauthoryear{Metcalf et~al.,}{Metcalf et~al.}{2019}]{Metcalf_2019}
Metcalf R.~B.,  et~al., 2019, \mn@doi [Astronomy \& Astrophysics] {10.1051/0004-6361/201832797}, 625, A119

\bibitem[\protect\citeauthoryear{{More}, {Cabanac}, {More}, {Alard}, {Limousin}, {Kneib}, {Gavazzi}  \& {Motta}}{{More} et~al.}{2012}]{2012ApJ...749...38M}
{More} A.,  {Cabanac} R.,  {More} S.,  {Alard} C.,  {Limousin} M.,  {Kneib} J.~P.,  {Gavazzi} R.,   {Motta} V.,  2012, \mn@doi [\apj] {10.1088/0004-637x/749/1/38}, \href {https://ui.adsabs.harvard.edu/abs/2012ApJ...749...38M} {749, 38}

\bibitem[\protect\citeauthoryear{O’Ryan et~al.,}{O’Ryan et~al.}{2023}]{O_Ryan_2023}
O’Ryan D.,  et~al., 2023, \mn@doi [ApJ] {10.3847/1538-4357/acc0ff}, 948, 40

\bibitem[\protect\citeauthoryear{Pearson}{Pearson}{1901}]{doi:10.1080/14786440109462720}
Pearson K.,  1901, \mn@doi [The London, Edinburgh, and Dublin Philosophical Magazine and Journal of Science] {10.1080/14786440109462720}, 2, 559

\bibitem[\protect\citeauthoryear{Pedregosa et~al.,}{Pedregosa et~al.}{2011}]{scikit-learn}
Pedregosa F.,  et~al., 2011, J. Mach. Learn. Res., 12, 2825

\bibitem[\protect\citeauthoryear{{Petrosian}}{{Petrosian}}{1976}]{1976ApJ...209L...1P}
{Petrosian} V.,  1976, \mn@doi [\apjl] {10.1086/182301}, \href {https://ui.adsabs.harvard.edu/abs/1976ApJ...209L...1P} {210, L53}

\bibitem[\protect\citeauthoryear{R., V., E.  \& M.}{R. et~al.}{1999}]{Barbon1999}
R. B.,  V. B.,  E. C.,   M. T.,  1999, \mn@doi [\aaps] {10.1051/aas:1999404}, \href {https://ui.adsabs.harvard.edu/abs/1999A&AS..139..531B} {139, 531}

\bibitem[\protect\citeauthoryear{Rasmussen \& Williams}{Rasmussen \& Williams}{2006}]{rasmussen2006gaussian}
Rasmussen C.~E.,  Williams C.~K.,  2006, Gaussian Processes for Machine learning.
MIT Press, Massachusetts, USA

\bibitem[\protect\citeauthoryear{Scaramella et~al.,}{Scaramella et~al.}{2022}]{Euclid_2022}
Scaramella R.,  et~al., 2022, \mn@doi [A&A] {10.1051/0004-6361/202141938}, 662, A112

\bibitem[\protect\citeauthoryear{{Shlens}}{{Shlens}}{2014}]{Shlens2014}
{Shlens} J.,  2014, \mn@doi [arXiv e-prints] {10.48550/arXiv.1404.1100}, \href {https://ui.adsabs.harvard.edu/abs/2014arXiv1404.1100S} {p. arXiv:1404.1100}

\bibitem[\protect\citeauthoryear{{Shu} et~al.,}{{Shu} et~al.}{2017}]{2017ApJ...851...48S}
{Shu} Y.,  et~al., 2017, \mn@doi [\apj] {10.3847/1538-4357/aa9794}, \href {https://ui.adsabs.harvard.edu/abs/2017ApJ...851...48S} {851, 48}

\bibitem[\protect\citeauthoryear{{Skrutskie} et~al.,}{{Skrutskie} et~al.}{2006}]{2006AJ....131.1163S}
{Skrutskie} M.~F.,  et~al., 2006, \mn@doi [\aj] {10.1086/498708}, \href {https://ui.adsabs.harvard.edu/abs/2006AJ....131.1163S} {131, 1163}

\bibitem[\protect\citeauthoryear{{Slijepcevic}, {Scaife}, {Walmsley}, {Bowles}, {Wong}, {Shabala}  \& {White}}{{Slijepcevic} et~al.}{2024}]{slijepcevic2023radio}
{Slijepcevic} I.~V.,  {Scaife} A. M.~M.,  {Walmsley} M.,  {Bowles} M.,  {Wong} O.~I.,  {Shabala} S.~S.,   {White} S.~V.,  2024, \mn@doi [RAS Techniques and Instruments] {10.1093/rasti/rzad055}, \href {https://ui.adsabs.harvard.edu/abs/2024RASTI...3...19S} {3, 19}

\bibitem[\protect\citeauthoryear{{Solarz}, {Bilicki}, {Gromadzki}, {Pollo}, {Durkalec}  \& {Wypych}}{{Solarz} et~al.}{2017}]{Solarz_2017}
{Solarz} A.,  {Bilicki} M.,  {Gromadzki} M.,  {Pollo} A.,  {Durkalec} A.,   {Wypych} M.,  2017, \mn@doi [\aap] {10.1051/0004-6361/201730968}, \href {https://ui.adsabs.harvard.edu/abs/2017A&A...606A..39S} {606, A39}

\bibitem[\protect\citeauthoryear{{Soroka}, {Meshcheryakov}  \& {Gerasimov}}{{Soroka} et~al.}{2022}]{MCAI}
{Soroka} A.,  {Meshcheryakov} A.,   {Gerasimov} S.,  2022, in {Ruiz} J.~E.,  {Pierfedereci} F.,   {Teuben} P.,  eds,  ASP Conf. Ser. Vol. 532, Astronomical Data Analysis Software and Systems XXX. Astron. Soc. Pac, San Francisco. p.~307 (\mn@eprint {arXiv} {2105.02958}), \mn@doi{10.48550/arXiv.2105.02958}

\bibitem[\protect\citeauthoryear{Sridhar et~al.,}{Sridhar et~al.}{2020}]{Sridhar_2020}
Sridhar S.,  et~al., 2020, \mn@doi [ApJ] {10.3847/1538-4357/abc0f0}, 904, 69

\bibitem[\protect\citeauthoryear{Storey-Fisher, Huertas-Company, Ramachandra, Lanusse, Leauthaud, Luo, Huang  \& Prochaska}{Storey-Fisher et~al.}{2021}]{Storey_Fisher_2021}
Storey-Fisher K.,  Huertas-Company M.,  Ramachandra N.,  Lanusse F.,  Leauthaud A.,  Luo Y.,  Huang S.,   Prochaska J.~X.,  2021, \mn@doi [MNRAS] {10.1093/mnras/stab2589}, 508, 2946

\bibitem[\protect\citeauthoryear{{Tan} \& {Le}}{{Tan} \& {Le}}{2020}]{tan2020efficientnet}
{Tan} M.,  {Le} Q.~V.,  2020, \mn@doi [arXiv e-prints] {10.48550/arXiv.1905.11946}, \href {https://ui.adsabs.harvard.edu/abs/2019arXiv190511946T} {p. arXiv:1905.11946}

\bibitem[\protect\citeauthoryear{{Taylor}}{{Taylor}}{2015}]{taylor2015topcats}
{Taylor} M.,  2015, \mn@doi [arXiv e-prints] {10.48550/arXiv.1512.06567}, \href {https://ui.adsabs.harvard.edu/abs/2015arXiv151206567T} {p. arXiv:1512.06567}

\bibitem[\protect\citeauthoryear{{The Astropy Collaboration}}{{The Astropy Collaboration}}{2013}]{refId0}
{The Astropy Collaboration} 2013, \mn@doi [A\&A] {10.1051/0004-6361/201322068}, 558, A33

\bibitem[\protect\citeauthoryear{{The Astropy Collaboration}}{{The Astropy Collaboration}}{2018}]{Price-Whelan_2018}
{The Astropy Collaboration} 2018, \mn@doi [AJ] {10.3847/1538-3881/aabc4f}, 156, 123

\bibitem[\protect\citeauthoryear{{The Astropy Collaboration}}{{The Astropy Collaboration}}{2022}]{TheAstropyCollaboration_2022}
{The Astropy Collaboration} 2022, \mn@doi [ApJ] {10.3847/1538-4357/ac7c74}, 935, 167

\bibitem[\protect\citeauthoryear{{The Dark Energy Survey Collaboration}}{{The Dark Energy Survey Collaboration}}{2005}]{https://doi.org/10.48550/arxiv.astro-ph/0510346}
{The Dark Energy Survey Collaboration} 2005, \mn@doi [arXiv e-prints] {10.48550/arXiv.astro-ph/0510346}, \href {https://ui.adsabs.harvard.edu/abs/2005astro.ph.10346T} {pp astro--ph/0510346}

\bibitem[\protect\citeauthoryear{{Toba} et~al.,}{{Toba} et~al.}{2014}]{2014ApJ...788...45T}
{Toba} Y.,  et~al., 2014, \mn@doi [\apj] {10.1088/0004-637x/788/1/45}, \href {https://ui.adsabs.harvard.edu/abs/2014ApJ...788...45T} {788, 45}

\bibitem[\protect\citeauthoryear{{Tutusaus}, {Sorce}, {Troja}  \& {Consortium}}{{Tutusaus} et~al.}{2023}]{tutusaus2023euclid}
{Tutusaus} I.,  {Sorce} J.,  {Troja} A.,   {Consortium} E.,  2023, in 41st International Conference on High Energy Physics. p.~95 (\mn@eprint {arXiv} {2211.08913}), \mn@doi{10.48550/arXiv.2211.08913}

\bibitem[\protect\citeauthoryear{{Vafaei Sadr}, {Bassett}  \& {Sekyi}}{{Vafaei Sadr} et~al.}{2022}]{sadr2022}
{Vafaei Sadr} A.,  {Bassett} B.~A.,   {Sekyi} E.,  2022, \mn@doi [arXiv e-prints] {10.48550/arXiv.2210.16334}, \href {https://ui.adsabs.harvard.edu/abs/2022arXiv221016334V} {p. arXiv:2210.16334}

\bibitem[\protect\citeauthoryear{Walmsley et~al.,}{Walmsley et~al.}{2019}]{10.1093/mnras/stz2816}
Walmsley M.,  et~al., 2019, \mn@doi [MNRAS] {10.1093/mnras/stz2816}, 491, 1554

\bibitem[\protect\citeauthoryear{Walmsley et~al.,}{Walmsley et~al.}{2021}]{Walmsley_2021}
Walmsley M.,  et~al., 2021, \mn@doi [MNRAS] {10.1093/mnras/stab2093}, 509, 3966

\bibitem[\protect\citeauthoryear{Walmsley et~al.,}{Walmsley et~al.}{2022}]{Walmsley_2022}
Walmsley M.,  et~al., 2022, \mn@doi [MNRAS] {10.1093/mnras/stac525}, 513, 1581

\bibitem[\protect\citeauthoryear{Walmsley et~al.,}{Walmsley et~al.}{2023}]{zoobot}
Walmsley M.,  et~al., 2023, \mn@doi [J. Open Source Softw.] {10.21105/joss.05312}, 8, 5312

\bibitem[\protect\citeauthoryear{Wenger et~al.,}{Wenger et~al.}{2000}]{Wenger_2000}
Wenger M.,  et~al., 2000, \mn@doi [Astronomy and Astrophysics Supplement Series] {10.1051/aas:2000332}, 143, 9

\bibitem[\protect\citeauthoryear{Wright et~al.,}{Wright et~al.}{2010}]{Wright_2010}
Wright E.~L.,  et~al., 2010, \mn@doi [AJ] {10.1088/0004-6256/140/6/1868}, 140, 1868

\bibitem[\protect\citeauthoryear{{York} et~al.,}{{York} et~al.}{2000}]{2000AJ....120.1579Y}
{York} D.~G.,  et~al., 2000, \mn@doi [\aj] {10.1086/301513}, \href {https://ui.adsabs.harvard.edu/abs/2000AJ....120.1579Y} {120, 1579}

\bibitem[\protect\citeauthoryear{{Zheng} \& {Shen}}{{Zheng} \& {Shen}}{2020}]{Zheng2020}
{Zheng} Y.-L.,  {Shen} S.-Y.,  2020, \mn@doi [\apjs] {10.3847/1538-4365/ab5c26}, \href {https://ui.adsabs.harvard.edu/abs/2020ApJS..246...12Z} {246, 12}

\bibitem[\protect\citeauthoryear{{de Vaucouleurs}}{{de Vaucouleurs}}{1948}]{1948AnAp...11..247D}
{de Vaucouleurs} G.,  1948, Ann. Astrophys., \href {https://ui.adsabs.harvard.edu/abs/1948AnAp...11..247D} {11, 247}

\bibitem[\protect\citeauthoryear{Ćiprijanović et~al.,}{Ćiprijanović et~al.}{2021}]{10.1093/mnras/stab1677}
Ćiprijanović A.,  et~al., 2021, \mn@doi [MNRAS] {10.1093/mnras/stab1677}, 506, 677

\makeatother
\end{thebibliography}

\appendix

\section{Computational Resources}
\label{Appendix: Computational Resources}
In this paper, our goal was to test \astronomaly{} on a large scale, which poses significant computational challenges. This appendix provides the details of the computational resources that we used for our analysis.
\subsection*{DECaLS\footnote{\url{https://www.legacysurvey.org/dr8/description/}} and storage}
The coadded image stack for DECaLS is stored in files totalling 45 TB of storage space, which is far too large to transfer and host locally. Cut-outs are used instead, as this allows greater freedom over the data selected and allows the user to determine how large each cut-out should be. With nearly 4 million sources, 150x150 pixels for each image, the storage space required for the data alone reaches nearly 150 GB. In addition, the catalogue, features, and other files created during the pipeline contribute an additional 100 GB needed. While not significant when it comes to storage, the transfer of such an amount does take a significant amount of time. Pre-processing several million sources incurs a significant amount of read-and-write operations to be done. If the images are processed and the output saved in a different location, the storage required is double the original amount. 

\subsection*{Computational times and other requirements}
Obtaining the data is straightforward with regard to accessibility, but the transference of data is extremely time-consuming. Obtaining the cut-outs of 3 884 404 sources took several weeks, with substantial failure rates and server maintenance impacting the process. The feature extraction process using the CNN is not very memory intensive but does use a large number of computational hours. The following lists the computational times and memory requirements for different steps of the pipeline. \\
\ \\
\textbf{Feature extraction:} It should be noted that a graphics processing unit (GPU) is \textit{highly} recommended for deep learning as it would significantly speed up the process. For this work though, only central processing units (CPUs) were used. The features file is almost 40GB in size (saved as a parquet format) and uses a similar amount of RAM to read in. This almost instantly rules out any local computing platform. \\
\ \\
\textit{CPU times: user 23h 1min 21s} \\
\ \\
\textbf{Dimensionality reduction:} Applying PCA on such a file, which has dimensions of 3 884 404 by 1280, is surprisingly quick using a CPU, but is memory intensive.\\
\ \\
\textit{CPU times: 2h 46min 30s \\ 
Max Memory: 228.64GB \\}
\ \\
\textbf{Isolation forest:} iForest uses smaller subsets of the input data set in an ensemble of decision trees which significantly reduces memory usage. Applied to the features which are reduced to a lower dimension with PCA: \\
\ \\
\textit{CPU times: 10min 17s \\
Max Memory: between 2 and 5Gb} \\
\ \\
However, if iForest is applied to the full features it is seen that the memory requirement increases significantly. \\
\ \\
\textit{CPU times: 1h 5min 25s \\
Max Memory: Oscillating between 38 and 91 GB} \\
\ \\
\textbf{Neighbour score:} \textit{Memory peaks at 11GB during retraining. Computational times vary between 20 and 30 min. \\}
\ \\
\textbf{Direct regression:} \textit{Failed at each attempt with memory usage exceeding the upper limit available at the time (232GB). \\
No computational time is known.}

\section{Data Selection Cuts}
\label{Appendix: Data Cuts}
The selection criteria in Section \ref{Data: Selection Cuts} is used to select a subset of the DECaLS data set and is done via TOPCat (Taylor \citeyear{taylor2015topcats}), using a Table Access Protocol (TAP) query. This type of query is written in the Astronomical Data Query Language (ADQL) as follows (Note the terms ``AND'' and ``OR'', referring to inclusive and exclusive data selection cuts, respectively):

\begin{lstlisting}[backgroundcolor = \color{lightgray}, language=SQL]
allmask_g=0 AND allmask_r=0 AND allmask_z=0 
rchisq_g>1 OR rchisq_r>1 OR rchisq_z>1
snr_g>0 AND snr_r>0 AND snr_z>0
flux_g>7 OR flux_r>7 OR flux_z>7
shapedev_r>3 OR shapeexp_r>3
\end{lstlisting}
\ \\
The DECaLS data are available with a tap query at \url{https://datalab.noirlab.edu/ls/dataAccess.php}. The following are more detailed descriptions of the various DECaLS catalogue entries incorporated in this work. \\
\ \\
\textbf{No point sources:} This value is used to remove all sources observed by the telescope that cannot be resolved. They would appear to be point sources and the inherent structure of the source is not identifiable. \\
\ \\
\textbf{ALLMASK}: These are masked sources, typically stellar in nature. Sources with a value greater than 0 in any of the three bands ($g$,$r$, and $z$) are not considered. \\
\ \\
\textbf{Reduced Chi-Square (rchi) values}: The `Profile-weighted $\chi^2$ of model fit normalized by the number of pixels in g, r, and z.' This serves as an indication of how well the model fits the source, although, as seen in Mao et al. \citeyear{Mao_2021}, the models do not fit as well as expected. Various values of this parameter were visually inspected using hundreds of images with the conclusion that the value in any one band should be at least 1; values lower than one are related to sources that do not have clearly visible structure. No upper limit restriction was placed because numerous sources were identified during the inspection that have high values for these parameters, but which are still well resolved. \\
\ \\
\textbf{SNR}: The standard signal-to-noise ratio is set to be greater than 0 in all bands. This is because sources with `negative' signal-to-noise ratios exist within the catalogue and are mainly artefacts. Using this restriction, a significant amount of artefacts are removed from the data sets. \\
\ \\
\textbf{Flux values}: The data set was further restricted by excluding sources with a flux value below 7 nanomaggies, as these sources are too dim to show any clear structure. This was observed from the visual inspection of the images. \\
\ \\
\textbf{Shapedev and shapeexp}: The `Half-light radius of de Vaucouleurs model ($>0$)' and `Half-light radius of exponential model ($>0$)', respectively. These are used to identify the image size to be used. Thousands of images of sources were inspected for various values of shapedev and shapeexp, with values ranging from 0 to 200. It was seen that for low values ($<3$), the sources are faint and it becomes difficult to identify any structure within them. \\

\bsp	
\label{lastpage}
\end{document}